\documentclass{revtex4-2}

\usepackage{hyperref}
\usepackage{amsfonts}
\usepackage{amsmath}
\usepackage{graphicx}
\usepackage{subcaption}
\usepackage{fancyhdr}
\usepackage[normalem]{ulem}

\hypersetup{
    colorlinks = true,
    citecolor = blue,
    urlcolor = blue
}

\newcommand{\p}[1]{\!\left( #1 \right)}

\newcommand{\bra}[1]{\left\langle #1 \right|}
\newcommand{\ket}[1]{\left| #1 \right\rangle}

\begin{document}

\title{Performance Analysis of Multi-Angle QAOA for \texorpdfstring{$p > 1$}{p greater than 1}}

\author{Igor Gaidai}
\email{igaidai@utk.edu}
\affiliation{Department of Industrial and Systems Engineering\\University of Tennessee at Knoxville\\Knoxville, TN 37996}
 
\author{Rebekah Herrman}
\affiliation{Department of Industrial and Systems Engineering\\University of Tennessee at Knoxville\\Knoxville, TN 37996}	

\begin{abstract}
In this paper we consider the scalability of Multi-Angle QAOA with respect to the number of QAOA layers. We found that MA-QAOA is able to significantly reduce the depth of QAOA circuits, by a factor of up to 4 for the considered data sets. However, MA-QAOA is not optimal for minimization of the total QPU time. Different optimization initialization strategies are considered and compared for both QAOA and MA-QAOA. Among them, a new initialization strategy is suggested for MA-QAOA that is able to consistently and significantly outperform random initialization used in the previous studies.
\end{abstract}

\maketitle

\section{Introduction}
The Quantum Approximate Optimization Algorithm (QAOA) \cite{Farhi2014, Choi2019, Blekos2023, hogg2000quantum} is a promising variational quantum algorithm designed to find approximate solutions to combinatorial optimization problems. With the appropriate transformations, \cite{Hadfield2019, Hadfield2021} the optimized cost function can be translated to the cost Hamiltonian $C$, and the original problem can be reformulated as finding the maximum eigenpair of $C$. 

The basic version of QAOA finds large eigenpairs of $C$ as follows. First, a second Hamiltonian $B$, called the mixing Hamiltonian, is introduced. It is usually defined as a sum of Pauli $X$ operators applied to each qubit,
\begin{gather}
    B = \sum_{i} X_i,
\end{gather}
although other mixers can be found in the literature \cite{bartschi2020grover, Wang2020, fuchs2022constrained, zhu2022adaptive}. Then an equal superposition state $\ket{+}^{\otimes n}$ (an eigenstate of $B$) is subjected to alternating evolution under the cost ($C$) and mixing ($B$) Hamiltonians for a total of $p$ layers.
\begin{gather}
    \ket{\gamma,\beta} = U\p{B, \beta_p} U\p{C, \gamma_p} ... U\p{B, \beta_1} U\p{C, \gamma_1} \ket{+}^{\otimes n} \\
    U\p{B, \beta} = e^{-i \beta B} \label{QAOAUB} \\
    U\p{C, \gamma} = e^{-i \gamma C} \label{QAOAUC}
\end{gather}
The total number of layers ($p$) is a convergence parameter of the algorithm and the individual evolution times of each Hamiltonian ($\gamma_1 ... \gamma_p$ $\beta_1 ... \beta_p$, also known as QAOA angles) are selected to maximize the expectation of the cost Hamiltonian in the final QAOA state $\bra{\gamma, \beta} C \ket{\gamma, \beta}$, usually via a classical optimizer. Finally, a series of repeated measurements of $\ket{\gamma,\beta}$ in the computational basis yields a bit string with a cost of at least $\bra{\gamma, \beta} C \ket{\gamma, \beta}$.

QAOA has been a subject of numerous recent studies (e.g., \cite{crooks2018performance, Wang2018, Wurtz2020, Basso2021, Marwaha2021, Basso2022, Boulebnane2022, Hadfield2022, herrman2022relating, Ozaeta2022, Lykov2023, Shaydulin2023, Stechly2023}), but many challenges still exist. One of such challenges is an efficient implementation of QAOA on the existing Noisy Intermediate Scale Quantum (NISQ) devices. NISQ devices suffer from high levels of noise,\cite{Sun2021, Dasgupta2021} which means that only circuits of limited depths\cite{herrman2021lower} can be executed before the qubits lose coherence and the result is completely dominated by noise. The number of QAOA layers necessary for efficient solution of practically relevant problems significantly exceeds the capacity of the modern quantum hardware. Recent work has studied methods for decomposing practical-sized problems to fit on hardware with fewer qubits, however some of these methods have drawbacks such as trying to piece together disjoint subproblem solutions \cite{zhou2023qaoa, ponce2023graph, li2022large}. Therefore, to make QAOA more useful on NISQ devices it is important to design a version of QAOA that minimizes the necessary number of layers.

One way to achieve this is to allow independent evolution times for each term of the cost and mixer Hamiltonians, which changes the definitions of Eqs. (\ref{QAOAUB}) and (\ref{QAOAUC}) to
\begin{gather}
    U\p{B, \vec{\beta}} = e^{-i \sum_i \beta_i X_i} = \prod_i e^{-i \beta_i X_i} \\
    U\p{C, \vec{\gamma}} = e^{-i \sum_i \gamma_i C_i} = \prod_i e^{-i \gamma_i C_i}
\end{gather}
Obviously, the performance of this modification cannot be worse than that of the original QAOA, since any QAOA result can be reproduced simply by setting all $\gamma_i = \gamma$ and all $\beta_i = \beta$, but how much better can it be? Does it justify the cost of having to optimize a significantly larger number of parameters?

This idea was originally suggested by Farhi et al., \cite{Farhi2017} who demonstrated optimistic, but very limited results. Later on, the performance of this modification, dubbed Multi-Angle QAOA (MA-QAOA), was further investigated on a larger data set and analyzed in \cite{Herrman2022} and \cite{Shi2022}, but the analysis was limited to $p = 1$. A similar recent development \cite{Vijendran2023} suggested to combine the ideas of MA-QAOA with an XY-mixer \cite{Wang2020} to achieve even better performance, but they also considered the case of $p = 1$ only. Other related works include additional parameters on well-defined subproblems \cite{wurtz2021classically} and adding problem-independent ansatz layers with multiple parameters \cite{chalupnik2022augmenting}.

In practice we would like to take advantage of as many layers as our hardware can support to achieve the desired performance level, which can extend beyond $p = 1$ even on NISQ devices. How many layers will we need for that and how well does MA-QAOA scale beyond $p = 1$? Can we realistically find good angles even as $p$ gets large? All of these questions need to be addressed to establish if MA-QAOA is useful for practical implementations on NISQ devices and the goal of this paper is to provide the answers to them.

\section{Results}

The performance of the methods considered in this paper is compared based on the approximation ratio that they can achieve for a given instance of the unweighted MaxCut problem, which is to partition the vertices of a graph into two subsets such that the number of edges between the sets is maximized.

The approximation ratio is defined in the usual way as the ratio of the largest found expectation of the cost Hamiltonian to the maximum achievable cut on a given graph (found by brute-forcing through all possible partitions):
\begin{gather}
    AR = \frac{\bra{\gamma, \beta} C \ket{\gamma, \beta}}{C_{max}}.
\end{gather}
The approximation ratio is a number in the range of $[0,1]$.

The graphs for the MaxCut problem were generated randomly with fixed number of nodes and edge probability (Erd\"os-R\'enyi model). In order to explore how the performance of the methods scales with respect to the graph characteristics we generated a total of 7 data sets with 1000 connected, non-isomorphic graphs of specific \textit{QAOA covering depth} in each (also referred to as \textit{c-depth}, for brevity), which is defined as a minimum number of QAOA layers necessary for it to see the whole graph (starting from any edge) \cite{Farhi2014}. This terminology is derived from \cite{farhi2020quantum}, where the authors discuss the minimum number of iterations $p$ that QAOA needs to ``cover" the entire graph. As such, it is expected to correlate with the performance of QAOA.

This metric is closely related to the more conventional graph diameter as
\begin{align}
   \mathrm{diameter} - 1 \le \mathrm{c{\text -}depth} \le \mathrm{diameter} + 1,
\end{align}
where the graph diameter is the maximum shortest distance between all pairs of vertices in the graph.
Therefore, graph diameter and c-depth can be regarded as approximately equal. The exact diameter distribution and other characteristics of our data sets are summarized in Table \ref{tab:set_characteristics}.

\begin{table}
    \centering
    \begin{tabular}{|c|c|c|c|c|c|c|}
        \hline
         Set \#& Nodes& Edge probability& c-depth& \#(diameter = c-depth - 1)& \#(diameter = c-depth)& \#(diameter = c-depth + 1)\\\hline
         1& 9& 0.6& 3& 663& 332& 5\\\hline
         2& 10& 0.6& 3& 762& 237& 1\\\hline
         3& 11& 0.6& 3& 799& 201& 0\\\hline
         4& 12& 0.6& 3& 823& 176& 1\\\hline
         5& 12& 0.2& 4& 32& 481& 487\\\hline
         6& 12& 0.1& 5& 7& 223& 770\\\hline
         7& 12& 0.1& 6& 12& 211& 777\\\hline
    \end{tabular}
    \caption{Characteristics of the seven data sets considered in this paper. The last three columns show the number of graphs with specific diameter within each data set.}
    \label{tab:set_characteristics}
\end{table}

The results of the quantum algorithms compared in this work are calculated on a classical state vector simulator. The source code of the simulator and the graph files are available at \cite{maqaoarepo}.

\subsection{Selection of angles for QAOA}
\label{sec:qaoa_strategies}

In general, it is difficult to find optimal QAOA parameters for the MaxCut problem analytically because the expected value function is a complex trigonometric function that depends on the degree of each vertex in the graph \cite{hadfield2018quantum}. Therefore, the optimal parameters are typically found by numerical optimization procedures. 

One of the biggest challenges with this approach is the problem of selecting good initial angles for such optimization procedures. Selecting random initial angles may require an exponential (in $p$) number of restarts to find the best angles due to a large number of local minima and barren plateaus in the optimization landscape of QAOA, especially for large problem sizes and at high depth \cite{lee2021parameters, wang2021noise}. As such, a large number of publications in the field has been devoted to finding heuristic strategies for selection of good initial angles \cite{Zhou2020, Boulebnane2021, Galda2021, Sack2021, Farhi2022, Sud2022, Sack2023, sureshbabu2023parameter}. 

For the sake of fair comparison with MA-QAOA, we wanted to use the best angles that we can find for QAOA within a reasonable budget of optimization attempts. Therefore, in this section we will compare empirical performance of several such heuristics reported in the literature and select the one that performs best. Specifically, the best heuristic is defined as the one that is able to exceed $\mathrm{AR} = 16 / 17 \approx 0.941$ for all graphs in a given data set, using the smallest number of QAOA  layers. The number $16 / 17$ was chosen because it is NP-hard to approximate MaxCut with $\mathrm{AR} > 16 / 17$, \cite{Vijendran2023, Hastad2001, Trevisan2000} therefore this level, in general, cannot be achieved with polynomial amount of resources (time) classically and it is interesting to see how much resources (number of layers) QAOA needs for the same task. 

\begin{figure}
    \centering
    \begin{subfigure}[b]{0.49\textwidth}
        \centering
        \includegraphics[width=\textwidth]{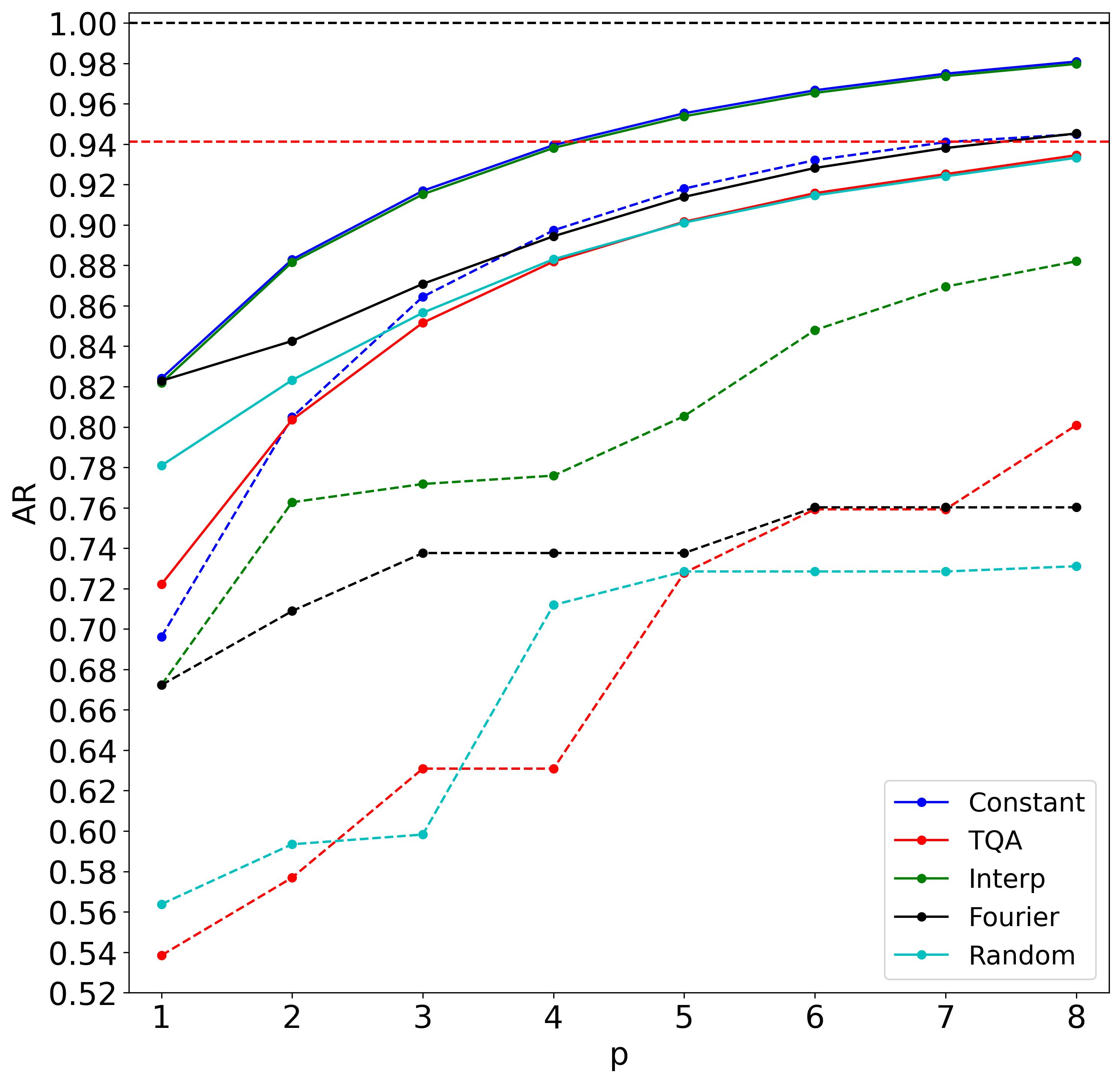}
        \caption{1 optimization at each level}
    \end{subfigure}
    \hfill
    \begin{subfigure}[b]{0.49\textwidth}
        \centering
        \includegraphics[width=\textwidth]{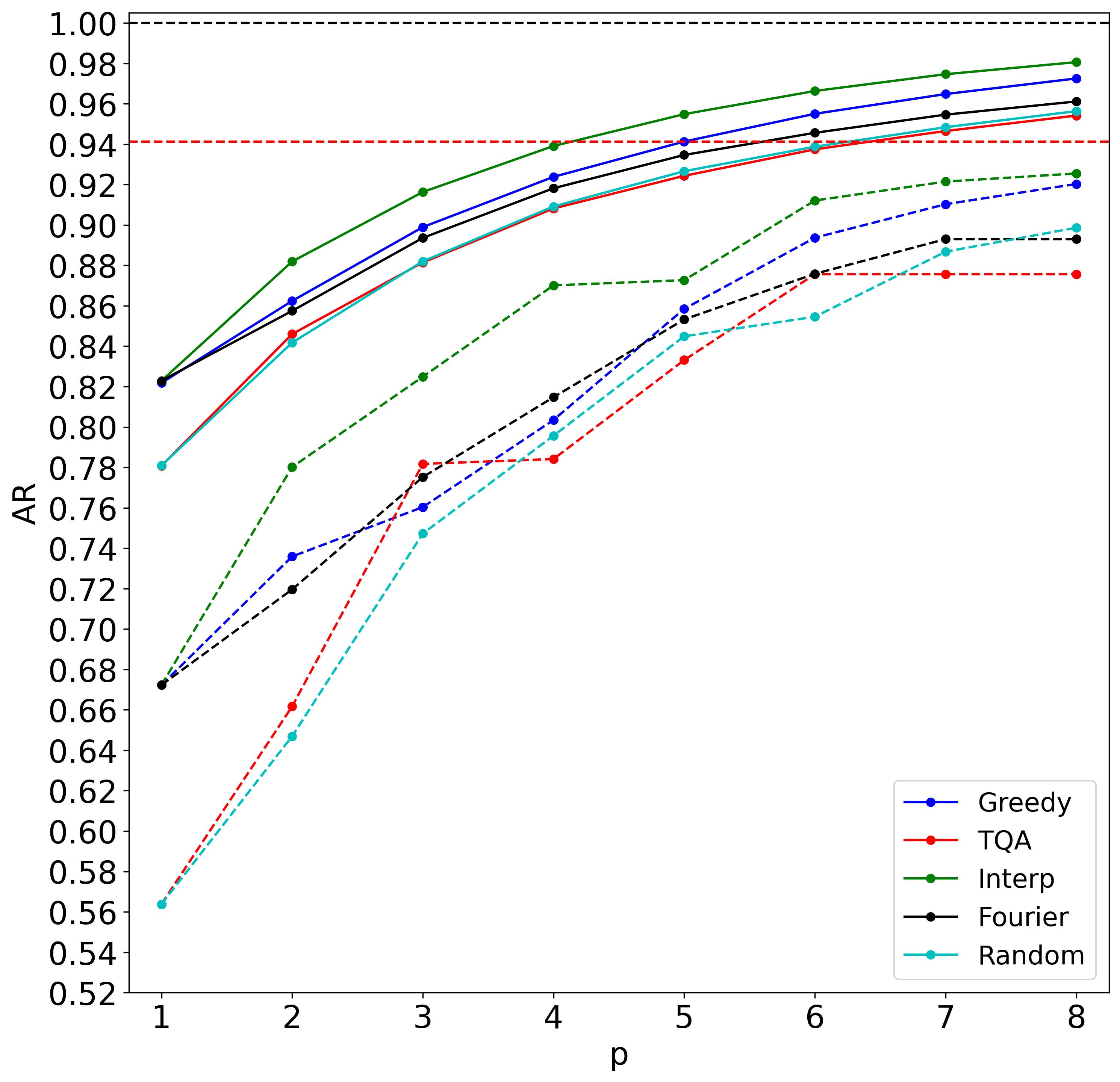}
        \caption{$p$ optimizations at level $p$}
    \end{subfigure}
    \caption{Comparison between different initialization heuristics for QAOA. Solid (dashed) lines show average (worst case) approximation ratio. The horizontal red dashed line shows the desired AR = 16/17.}
    \label{fig:qaoa-heuristics}
\end{figure}

The heuristics are compared on the data set with 9 nodes (the first row of Table \ref{tab:set_characteristics}). Each method was used to calculate all values of $p$ consecutively, until one of them exceeds $\mathrm{AR} = 16 / 17$ on every single graph in every data set. If a method failed to find better angles at a given level $p$, then the best angles found on level $p - 1$ are appended with zeros, and are declared as the best at level $p$. Therefore, the performance of each method can only increase monotonically with $p$. The average and worst case performances of the considered methods are shown in Figure \ref{fig:qaoa-heuristics}. The left frame shows the case when only 1 optimization attempt was allowed for each value of $p$. On the right frame, the methods were allowed to perform $p$ optimizations at a given value of $p$.

The following heuristics were considered: Constant \cite{Boulebnane2022}, TQA \cite{Sack2021}, Interp \cite{Zhou2020}, Fourier \cite{Zhou2020}, Greedy \cite{Sack2023}, Random. 

The first considered heuristic, labeled as ``Constant" here, was taken from Ref. \cite{Boulebnane2022}. There, the authors suggest to initialize all values of $\gamma$ with -0.01 and all values of $\beta$ with 0.01, i.e. a fixed constant that does not depend on anything and has opposite signs on $\gamma$ and $\beta$. We examined convergence with constants in the set $\{0.01, 0.05, 0.1, 0.2, 0.4, 1\}$ (See Supplemental Information for details) and found that QAOA converged fastest when the value was set to 0.2 and appeared to decrease with the distance away from 0.2. This constant was used to generate the corresponding line in Figure \ref{fig:qaoa-heuristics}. This heuristic is only plotted in the left frame of Figure \ref{fig:qaoa-heuristics}, since it cannot make use of more than 1 optimization attempt.

The next heuristic, labeled as ``TQA" (Trotterized Quantum Annealing), \cite{Sack2021} assumes, as the name suggests, that good starting angles can be obtained by using linear parameter schedules typical for quantum annealing. Specifically, the values of $\gamma$ are given by $\gamma_i = (i - 0.5)\Delta t / p$, and the values of $\beta$ are given by $\beta = \Delta t - \gamma$, where $\Delta t$ is the only unknown parameter, found by optimization in the one-dimensional $\Delta t$-space, after which the angles corresponding to the optimal $\Delta t$ are used as a starting point for the full $2p$-dimensional QAOA optimization. In the case of $p$ optimization attempts (the right frame of Figure \ref{fig:qaoa-heuristics}), the first one-dimensional optimization was not counted in the optimization budget as a separate optimization. Additionally, since the number of local minima in the one-dimensional space can be rather limited, if the optimization procedure returned the value of $\Delta t$ that has already been tried, then a random new value of $\Delta t$ was selected. In our experience, we observed that starting the full optimization from a random non-optimal value of $\Delta t$ can often perform better than starting from the optimal one.

The next two heuristics are taken from Ref. \cite{Zhou2020} and labeled as ``Interp" and ``Fourier". These heuristics are similar in that they assume that the optimal angles change smoothly and re-evaluate the best angles found at level $p - 1$ on a denser grid or use Fourier transform to generate a good starting guess for level $p$. The extra optimization attempts on right frame of Figure \ref{fig:qaoa-heuristics} are used to apply random perturbations to the best angles found at level $p - 1$ and start the optimization from there, in accordance with the prescription of Ref. \cite{Zhou2020}.

Another heuristic, labeled ``Greedy" can be found on the right frame of Figure \ref{fig:qaoa-heuristics} only, and replaces ``Constant" from the left frame. This heuristic (introduced in \cite{Sack2023}) is based on the idea that one can always insert an ``empty" QAOA layer (i.e. a layer with $\gamma_i = \beta_i = 0$) before or after any other QAOA layer, which will have no effect on the expectation value of the cost Hamiltonian. As such, these zero insertions to the angle vectors can be used to generate good initial guesses for level $p$, using the best angles found at level $p-1$. Indeed, starting the optimization from such angles guarantees that the expectation value achieved at level $p$ will be no worse than the one achieved at level $p-1$. There are $p$ possible choices for the insertion position to angles found at level $p-1$, therefore $p$ optimizations are necessary to explore all initial guesses obtained in this way, which, in fact, determined the choice of the optimization budget for the right frame of Figure \ref{fig:qaoa-heuristics}.

Finally, the last considered method is to select the angles randomly, which defines the lowest reasonable performance threshold for any method. This method was used as an experimental control against which to compare the other initialization heuristics.

For all heuristics, except Greedy, L-BFGS-B optimization method from scipy.optimize.minimize python package was used. For Greedy, we used Nelder-Mead, since BFGS is a first-order method and is not suitable for optimization from a transition state, where the first-order derivatives are zero. For heuristics that require the best angles from the previous layer (Interp, Fourier and Greedy), at $p = 1$, we used a random initialization with 10 attempts (in both frames of Figure \ref{fig:qaoa-heuristics}) to start them with globally optimal angles.

As one can see from Figure \ref{fig:qaoa-heuristics}, Constant demonstrates superior average and worst-case performance for all $p$, even compared to methods that were allowed $p$ optimization attempts. Interp has nearly the same performance on average, but is less consistent and does not do as well in the worst case even with added perturbations (on the right frame). Constant was the first to reach the NP-hard region of $\mathrm{AR} > 16/17$ in the worst case at $p = 8$ and thus, Constant was selected to represent the performance of QAOA in the comparison against MA-QAOA (Section~\ref{sec:comparison}).

\subsection{Selection of angles for MA-QAOA}
\label{sec:ma}

In the previous MA-QAOA studies \cite{Farhi2017, Herrman2022, Shi2022} the initial angles were selected randomly. However, as seen in the Section~\ref{sec:qaoa_strategies}, other techniques for finding initial angles tend to significantly outperform random initialization. Is it possible to adapt some of the QAOA heuristics to MA-QAOA? Will they still work well? In this section we introduce several new angle strategies for MA-QAOA and answer these questions.

The primary candidates for adaptation are the ones that perform best for QAOA. Looking at Figure \ref{fig:qaoa-heuristics}, we select Constant and Interp as such. Constant applies to MA-QAOA straightforwardly (set all $\gamma_i = 0.2$, and all $\beta_i = -0.2$). For Interp, we interpolate each angle $\gamma_i$ and $\beta_i$ separately, in the same way as it was for QAOA.

Another thing that we could try for MA-QAOA is to start the optimization from the optimal angles found by QAOA (with the Constant strategy, as discussed in the previous section), which we will call ``QAOA Relax".

To determine the role of quality of angles in QAOA Relax strategy, we also consider ``Random QAOA", where the initial values of $\gamma_i$ and $\beta_i$ are random, but equal within the same layer, as in QAOA. Finally, the last strategy is a completely random independent guess, which was the approach used in the previous studies \cite{Farhi2017, Herrman2022, Shi2022}.

\begin{figure}
    \centering
    \includegraphics[width=0.5\textwidth]{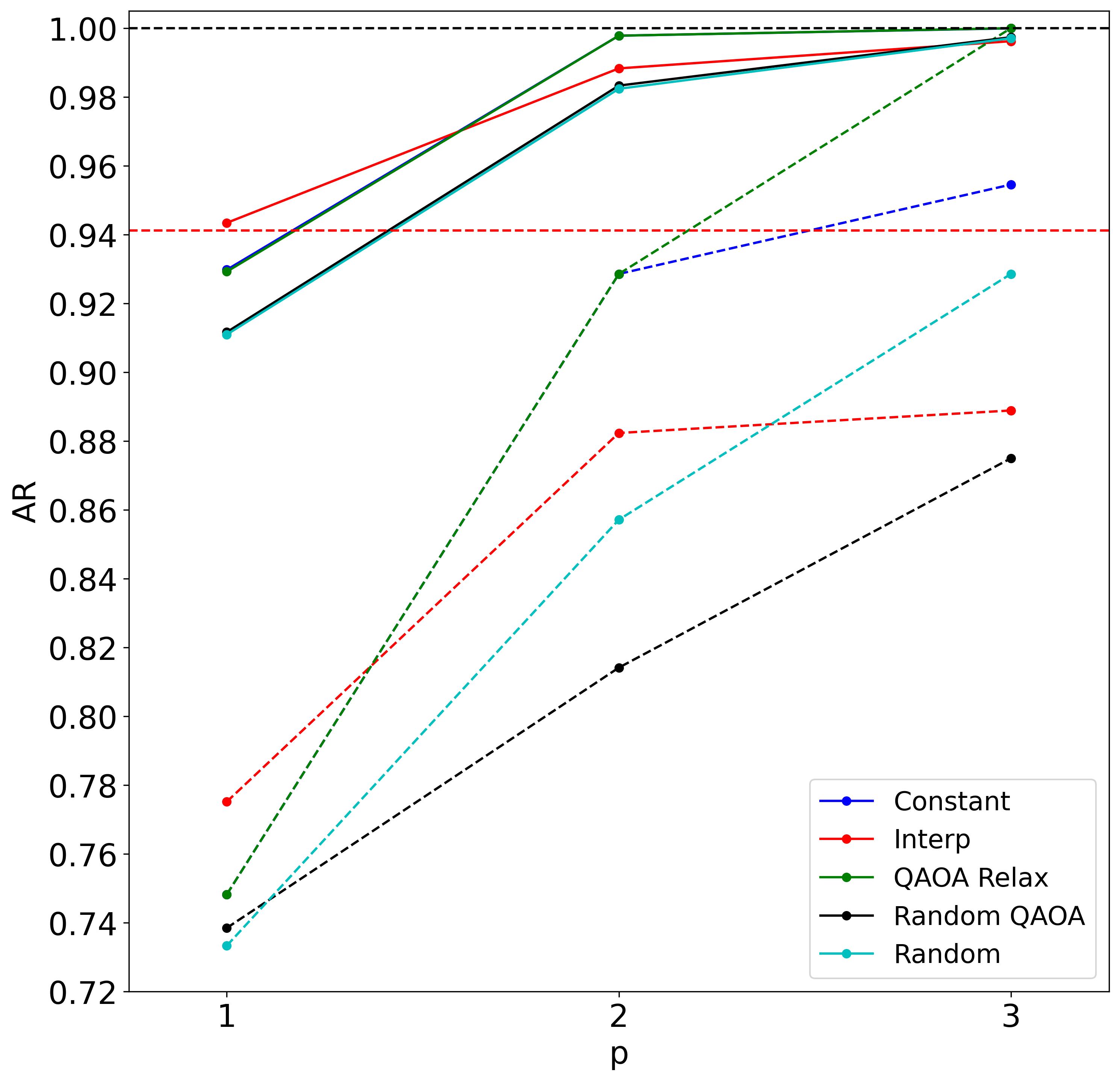}
    \caption{Comparison between different initialization heuristics for MA-QAOA on the data set with 9 nodes. Solid (dashed) lines show average (worst case) approximation ratio. The horizontal red dashed line shows the desired AR = 16/17.}
    \label{fig:ma-heuristics}
\end{figure}

The results of these strategies are shown in Figure \ref{fig:ma-heuristics} on the nine-vertex data set. Random QAOA and Random strategies show nearly the same performance on average, both worse than QAOA Relax. This indicates that simply keeping the angles the same within each layer does not improve performance and that the quality of QAOA angles for the QAOA Relax strategy is important to achieve quick convergence.

In contrast to QAOA, Interp does not perform well as an MA-QAOA initialization approach. In fact, the AR with the Interp initialization strategy gets even worse than the Random initialization strategy by $p = 3$, which indicates that optimal MA-QAOA angles tend to not change smoothly.

Both Constant and QAOA Relax are nearly identical on average and both reach the target convergence at $p = 3$, but QAOA Relax performs better in the worst case and is able to solve all MaxCut problems in the data set exactly. As such, we select it as the best strategy for comparison with QAOA below.

\subsection{Performance comparison between QAOA and MA-QAOA}
\label{sec:comparison}

\begin{figure}
    \centering
    \begin{subfigure}[b]{0.49\textwidth}
        \centering
        \includegraphics[width=\textwidth]{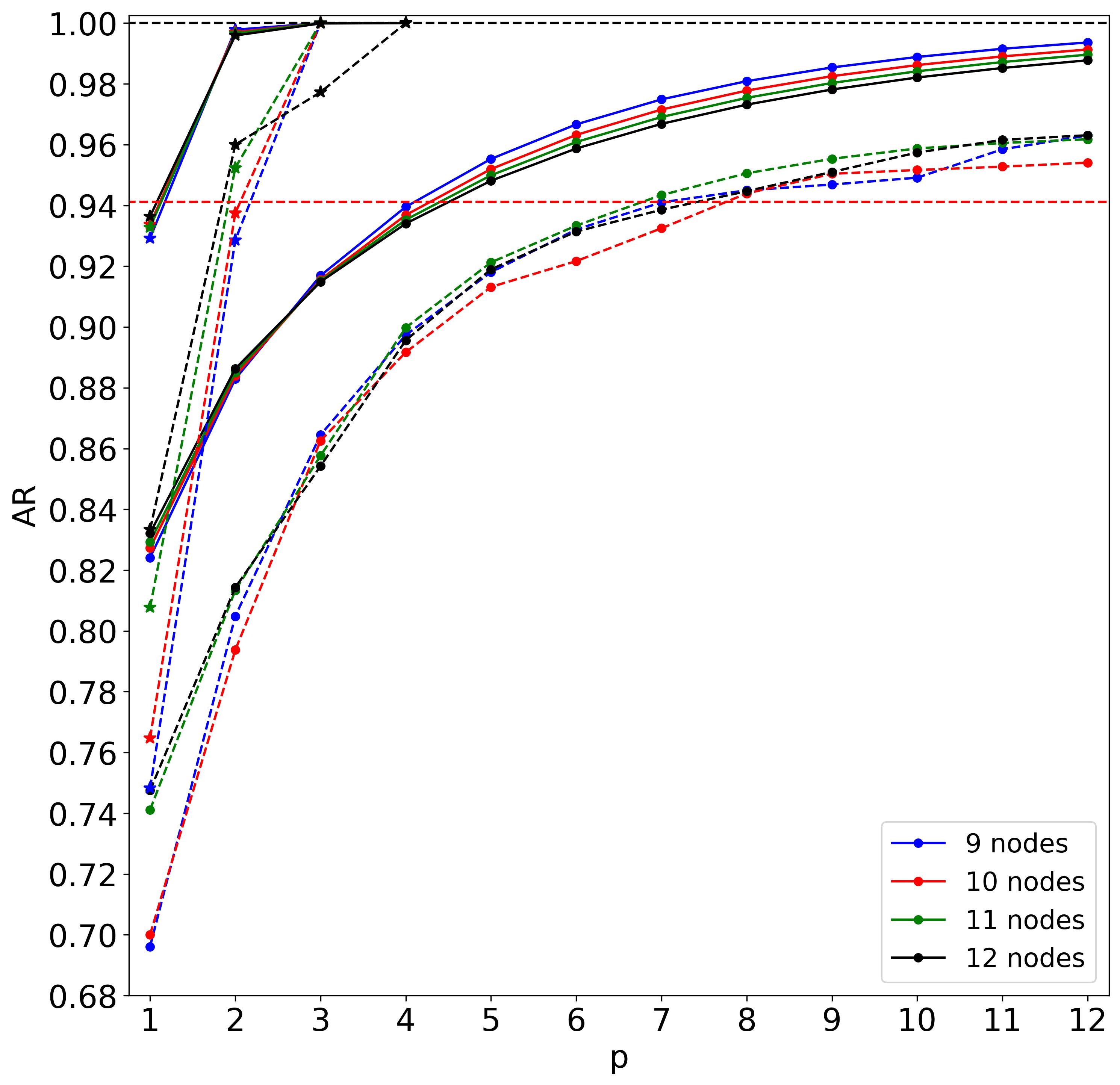}
        \caption{Fixed c-depth = 3, varying number of nodes}\label{fig:qaoa-ma-performance_1}
    \end{subfigure}
    \hfill
    \begin{subfigure}[b]{0.49\textwidth}
        \centering
        \includegraphics[width=\textwidth]{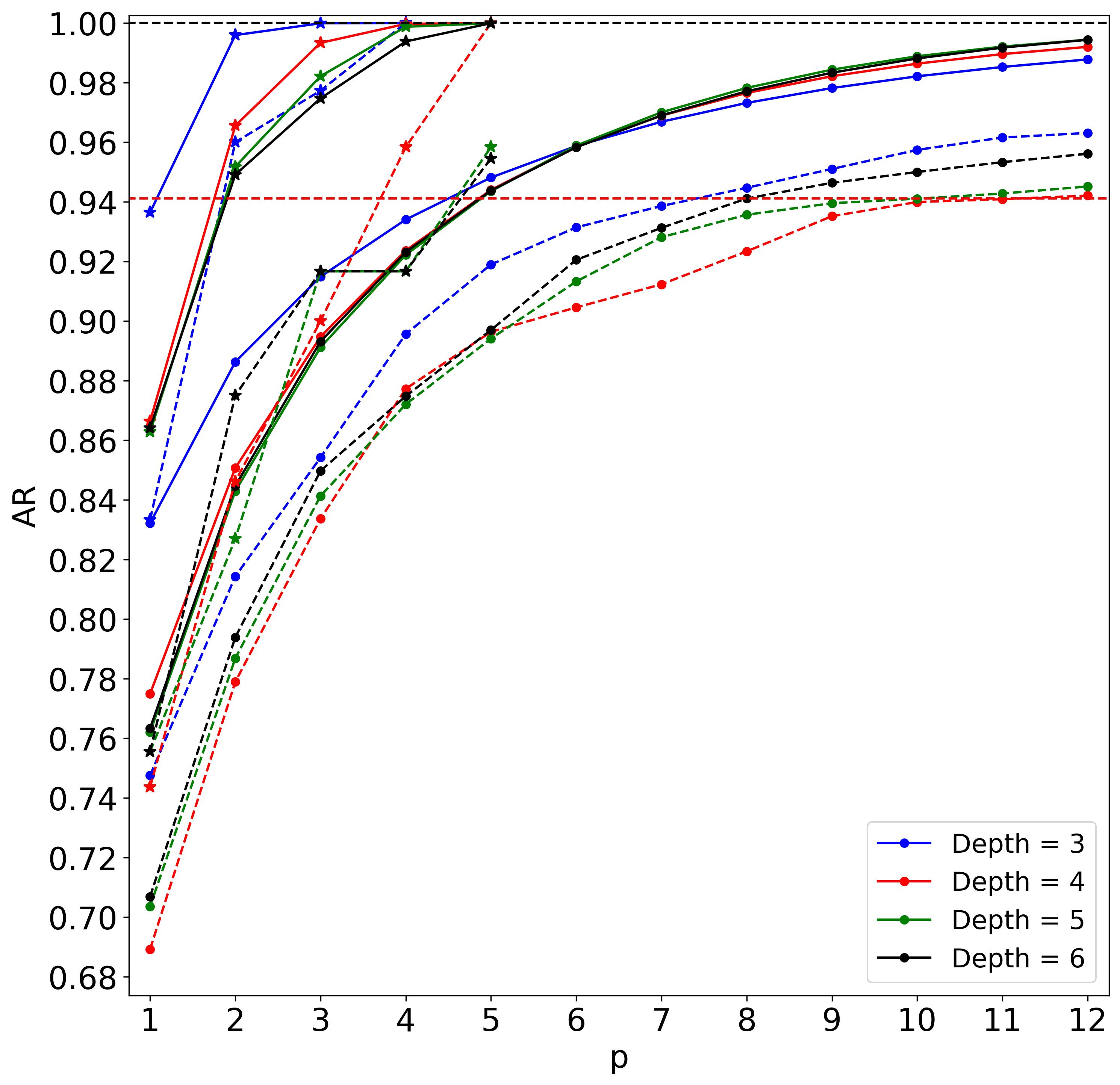}
        \caption{Fixed number of nodes = 12, varying c-depth}\label{fig:qaoa-ma-performance_2}
    \end{subfigure}
    \caption{Approximation ratio as a function of $p$ obtained with the best angle selection strategies for QAOA and MA-QAOA. Solid (dashed) lines show average (worst case) approximation ratio. Circles (stars) show the results of QAOA (MA-QAOA). The horizontal red dashed line shows the desired AR = 16/17.}
    \label{fig:qaoa-ma-performance}
\end{figure}

Using the best angle strategies picked in the two previous sections (Constant for QAOA and QAOA Relax for MA-QAOA), we calculated the AR on all data sets from Table \ref{tab:set_characteristics}. The results of this are shown in Figure \ref{fig:qaoa-ma-performance}. As one can see, the average AR of both QAOA and MA-QAOA converges to 1 smoothly. The worst case AR is generally less smooth, but still follows a similar trend. At sufficiently large values of $p$, the performance of both QAOA and MA-QAOA expectedly drops as the system size is increasing, but MA-QAOA is less affected by it. Note, that at low values of $p$, the order of the lines in Figure \ref{fig:qaoa-ma-performance_1} is reversed, which indicates that the performance of QAOA-like methods in general cannot be adequately judged based on the results for $p = 1$ only, and convergence with respect to $p$ has to be analyzed.

Interestingly, from Figure \ref{fig:qaoa-ma-performance_2}, one can see that QAOA converges faster for graphs of larger c-depth, while for MA-QAOA the behavior is the opposite. This might prompt the conclusion that QAOA could become better than MA-QAOA for graphs of large c-depth. Note, however, that the maximum c-depth is limited by the number of nodes in a graph and increasing the number of nodes has the opposite and stronger effect on the performance of QAOA. Additionally, the performance of MA-QAOA cannot be worse than performance of QAOA since we start the optimization from the optimal QAOA angles.

In both frames of Figure \ref{fig:qaoa-ma-performance}, we see that MA-QAOA achieves substantially larger ARs and thus converges much faster than QAOA. In particular, the number of layers necessary to converge the average AR required by QAOA was 5 for all data sets, while MA-QAOA was able to achieve the same performance level in 2 layers, reducing the depth of the circuit by a factor of 2.5. The number of layers necessary for the worst case AR is more variable and is given in Table \ref{tab:num_layers}. As one can see, in this case MA-QAOA allows to reduce the depth of the circuit by a factor of up to 4. As a consequence of the trends described above, the depth reduction factors tend to increase with the graph size (sets 1-4), but decrease with c-depth (sets 4-7).

\begin{table}
    \centering
    \begin{tabular}{|c|c|c|c|}
        \hline
         Set \#& \# layers QAOA& \# layers MA-QAOA& Depth reduction factor\\\hline
         1& 8& 3& 2.7\\\hline
         2& 8& 3& 2.7\\\hline
         3& 7& 2& 3.5\\\hline
         4& 8& 2& 4.0\\\hline
         5& 12& 4& 3.0\\\hline
         6& 11& 5& 2.2\\\hline
         7& 9& 5& 1.8\\\hline
    \end{tabular}
    \caption{The number of layers necessary to converge the worst-case approximation ratios for QAOA and MA-QAOA. The set numbers are the same as in Table \ref{tab:set_characteristics}.}
    \label{tab:num_layers}
\end{table}

\subsection{Prospects of MA-QAOA in fault-tolerant era}
\label{sec:cost}

In the previous section we demonstrated that the depth of QAOA circuit can be made several times smaller by using MA-QAOA. But what if we have a hypothetical fault-tolerant computer and the depth does not matter as much anymore? In that case the new efficiency metric that one can define instead of circuit depth, is the total time it takes to solve the problem, which can be approximated as the number of calls to QPU times depth of the corresponding circuits. Both QAOA and MA-QAOA have the same depth per layer, therefore for the sake of comparison between them, one can measure circuit depth simply in the number of QAOA layers.
\begin{equation}
    \mathrm{Cost} = n_c * p
    \label{eq:cost}
\end{equation}
where $n_c$ symbolizes the number of calls to a QPU.

\begin{figure}
    \centering
    \begin{subfigure}[b]{0.49\textwidth}
        \centering
        \includegraphics[width=\textwidth]{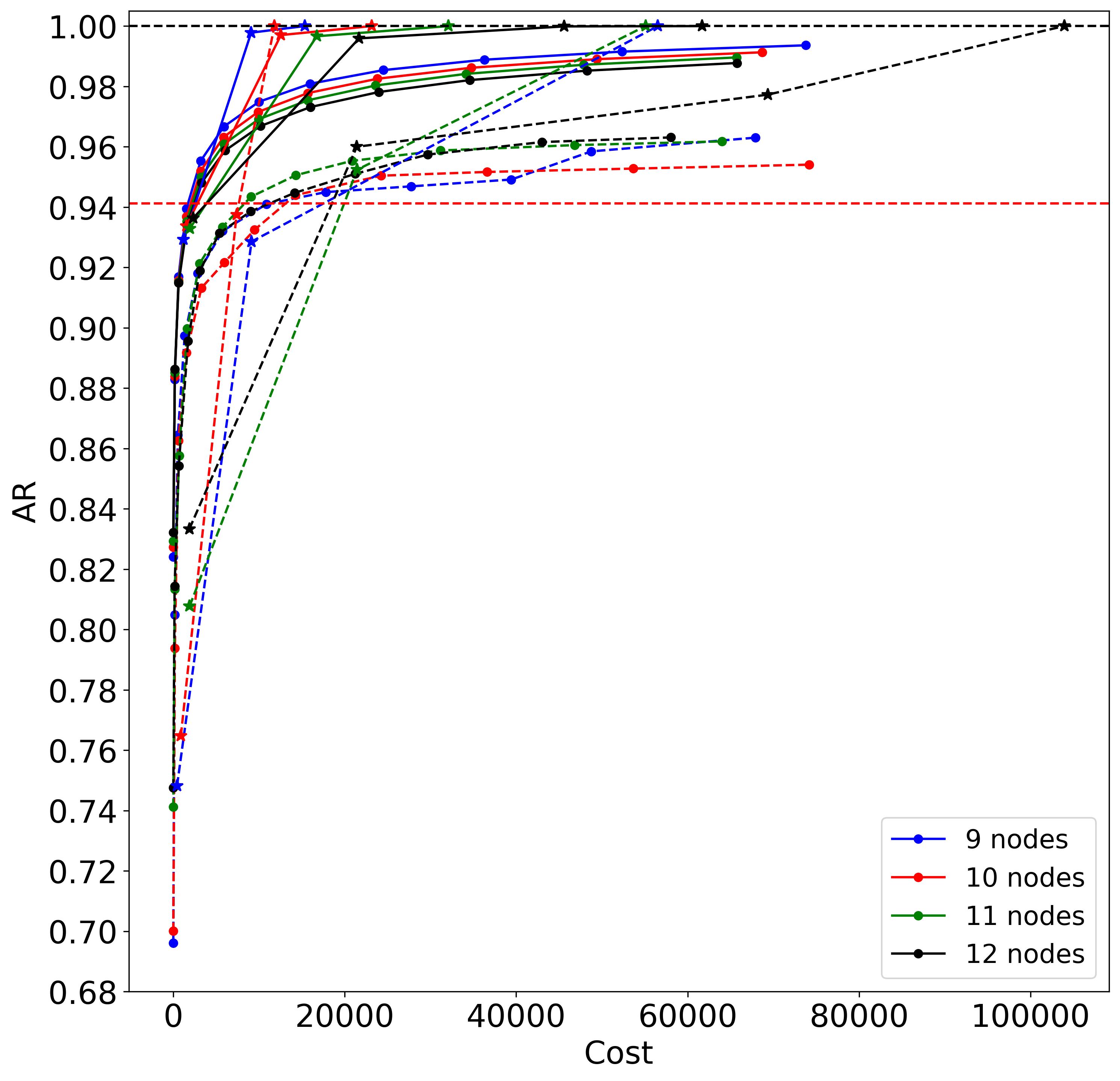}
        \caption{Fixed c-depth = 3, varying number of nodes}
    \end{subfigure}
    \hfill
    \begin{subfigure}[b]{0.49\textwidth}
        \centering
        \includegraphics[width=\textwidth]{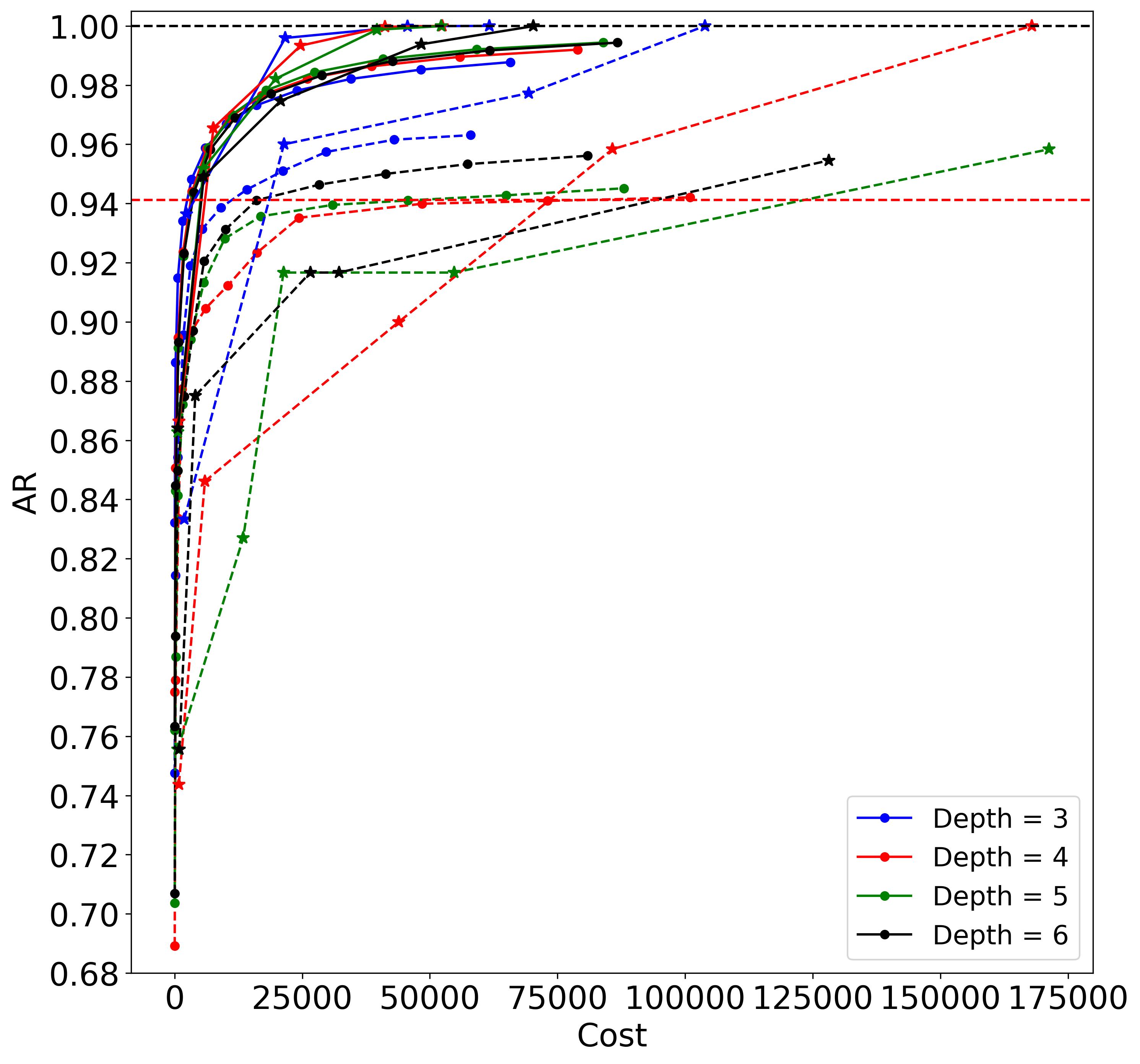}
        \caption{Fixed number of nodes = 12, varying c-depth}
    \end{subfigure}
    \caption{The data of Figure \ref{fig:qaoa-ma-performance}, re-plotted using the cost defined in Eq. \ref{eq:cost}.}
    \label{fig:qaoa-ma-performance-cost}
\end{figure}

Higher dimensionality of the optimization space means more calls to QPU. Will MA-QAOA still be better than QAOA with this metric, i.e. will it have better AR for a given cost? In Figure \ref{fig:qaoa-ma-performance-cost}, we re-plotted the data of Figure \ref{fig:qaoa-ma-performance}, using the metric of Eq. \ref{eq:cost} as the argument. As one can see, overall, QAOA and MA-QAOA are closely matched in the region of smaller values of $p$. On average, QAOA achieves the desired accuracy level at a substantially smaller cost compared to MA-QAOA. The same is true for most (5 out of 7) data sets in the worst case. However, starting at the sufficiently high values of $p$ ($p > 1$, data set dependent), MA-QAOA is able to achieve consistently larger ARs at any given cost. Considering the scaling with respect to the system size, the minimum value of $p$ at which this happens is expected to grow.

\section{Conclusions}
In this work, we examined multiple parameter setting strategies for QAOA (Constant, TQA, Interp, Fourier, Greedy, Random) and MA-QAOA (Constant, Interp, QAOA Relax, Random QAOA, Random), and determined the necessary number of QAOA layers for convergence on several MaxCut data sets. The key takeaways from the numerical experiments are:

\begin{enumerate}
    \item Setting the initial angles to a fixed constant with opposite signs for $\gamma$ and $\beta$ was the most efficient optimization initialization strategy for MaxCut QAOA among the strategies considered here.
    \item The same strategy also works well for MA-QAOA, but in the worst case it is better to start the optimization from the optimal QAOA angles instead.
    \item MA-QAOA allows to significantly reduce the depth of QAOA circuit (1.8 to 4 times for the data sets considered here), therefore MA-QAOA is well-suited for practical implementations of QAOA on NISQ devices.
    \item In cases when total QPU time needs to be minimized instead of depth (e.g. on a hypothetical fault-tolerant quantum computer), MA-QAOA loses to QAOA.
\end{enumerate}

As noted in the work, some QAOA angle initialization strategies do not work well for MA-QAOA. It is reasonable to assume that there may be MA-QAOA parameter initialization strategies that are not well-defined for QAOA due to the drastic difference in the number of parameters the algorithms require. Future work includes developing new angle initialization techniques for MA-QAOA that may not necessarily make sense for QAOA. Another research direction is to extend these tests to data sets with larger number of nodes ($> 100$), where the MaxCut problem may become challenging for the classical methods and see if the depth reduction achieved by MA-QAOA will be sufficient to demonstrate quantum advantage on NISQ devices.

\acknowledgments

R.\ Herrman acknowledges the National Science Foundation award CCF-2301120.

\bibliography{references}

\begin{thebibliography}{50}%
\makeatletter
\providecommand \@ifxundefined [1]{%
 \@ifx{#1\undefined}
}%
\providecommand \@ifnum [1]{%
 \ifnum #1\expandafter \@firstoftwo
 \else \expandafter \@secondoftwo
 \fi
}%
\providecommand \@ifx [1]{%
 \ifx #1\expandafter \@firstoftwo
 \else \expandafter \@secondoftwo
 \fi
}%
\providecommand \natexlab [1]{#1}%
\providecommand \enquote  [1]{``#1''}%
\providecommand \bibnamefont  [1]{#1}%
\providecommand \bibfnamefont [1]{#1}%
\providecommand \citenamefont [1]{#1}%
\providecommand \href@noop [0]{\@secondoftwo}%
\providecommand \href [0]{\begingroup \@sanitize@url \@href}%
\providecommand \@href[1]{\@@startlink{#1}\@@href}%
\providecommand \@@href[1]{\endgroup#1\@@endlink}%
\providecommand \@sanitize@url [0]{\catcode `\\12\catcode `\$12\catcode `\&12\catcode `\#12\catcode `\^12\catcode `\_12\catcode `\%12\relax}%
\providecommand \@@startlink[1]{}%
\providecommand \@@endlink[0]{}%
\providecommand \url  [0]{\begingroup\@sanitize@url \@url }%
\providecommand \@url [1]{\endgroup\@href {#1}{\urlprefix }}%
\providecommand \urlprefix  [0]{URL }%
\providecommand \Eprint [0]{\href }%
\providecommand \doibase [0]{https://doi.org/}%
\providecommand \selectlanguage [0]{\@gobble}%
\providecommand \bibinfo  [0]{\@secondoftwo}%
\providecommand \bibfield  [0]{\@secondoftwo}%
\providecommand \translation [1]{[#1]}%
\providecommand \BibitemOpen [0]{}%
\providecommand \bibitemStop [0]{}%
\providecommand \bibitemNoStop [0]{.\EOS\space}%
\providecommand \EOS [0]{\spacefactor3000\relax}%
\providecommand \BibitemShut  [1]{\csname bibitem#1\endcsname}%
\let\auto@bib@innerbib\@empty
\bibitem [{\citenamefont {Farhi}\ \emph {et~al.}(2014)\citenamefont {Farhi}, \citenamefont {Goldstone},\ and\ \citenamefont {Gutmann}}]{Farhi2014}%
  \BibitemOpen
  \bibfield  {author} {\bibinfo {author} {\bibfnamefont {E.}~\bibnamefont {Farhi}}, \bibinfo {author} {\bibfnamefont {J.}~\bibnamefont {Goldstone}},\ and\ \bibinfo {author} {\bibfnamefont {S.}~\bibnamefont {Gutmann}},\ }\href@noop {} {\bibinfo {title} {A quantum approximate optimization algorithm}} (\bibinfo {year} {2014}),\ \Eprint {https://arxiv.org/abs/1411.4028} {arXiv:1411.4028} \BibitemShut {NoStop}%
\bibitem [{\citenamefont {Choi}\ and\ \citenamefont {Kim}(2019)}]{Choi2019}%
  \BibitemOpen
  \bibfield  {author} {\bibinfo {author} {\bibfnamefont {J.}~\bibnamefont {Choi}}\ and\ \bibinfo {author} {\bibfnamefont {J.}~\bibnamefont {Kim}},\ }\bibfield  {title} {\bibinfo {title} {A tutorial on quantum approximate optimization algorithm (qaoa): Fundamentals and applications},\ }in\ \href {https://doi.org/10.1109/ICTC46691.2019.8939749} {\emph {\bibinfo {booktitle} {2019 International Conference on Information and Communication Technology Convergence (ICTC)}}}\ (\bibinfo {year} {2019})\ pp.\ \bibinfo {pages} {138--142}\BibitemShut {NoStop}%
\bibitem [{\citenamefont {Blekos}\ \emph {et~al.}(2023)\citenamefont {Blekos}, \citenamefont {Brand}, \citenamefont {Ceschini}, \citenamefont {Chou}, \citenamefont {Li}, \citenamefont {Pandya},\ and\ \citenamefont {Summer}}]{Blekos2023}%
  \BibitemOpen
  \bibfield  {author} {\bibinfo {author} {\bibfnamefont {K.}~\bibnamefont {Blekos}}, \bibinfo {author} {\bibfnamefont {D.}~\bibnamefont {Brand}}, \bibinfo {author} {\bibfnamefont {A.}~\bibnamefont {Ceschini}}, \bibinfo {author} {\bibfnamefont {C.-H.}\ \bibnamefont {Chou}}, \bibinfo {author} {\bibfnamefont {R.-H.}\ \bibnamefont {Li}}, \bibinfo {author} {\bibfnamefont {K.}~\bibnamefont {Pandya}},\ and\ \bibinfo {author} {\bibfnamefont {A.}~\bibnamefont {Summer}},\ }\href@noop {} {\bibinfo {title} {A review on quantum approximate optimization algorithm and its variants}} (\bibinfo {year} {2023}),\ \Eprint {https://arxiv.org/abs/2306.09198} {arXiv:2306.09198} \BibitemShut {NoStop}%
\bibitem [{\citenamefont {Hogg}(2000)}]{hogg2000quantum}%
  \BibitemOpen
  \bibfield  {author} {\bibinfo {author} {\bibfnamefont {T.}~\bibnamefont {Hogg}},\ }\bibfield  {title} {\bibinfo {title} {Quantum search heuristics},\ }\href {https://doi.org/https://doi.org/10.1103/PhysRevA.61.052311} {\bibfield  {journal} {\bibinfo  {journal} {Physical Review A}\ }\textbf {\bibinfo {volume} {61}},\ \bibinfo {pages} {052311} (\bibinfo {year} {2000})}\BibitemShut {NoStop}%
\bibitem [{\citenamefont {Hadfield}\ \emph {et~al.}(2019)\citenamefont {Hadfield}, \citenamefont {Wang}, \citenamefont {O'Gorman}, \citenamefont {Rieffel}, \citenamefont {Venturelli},\ and\ \citenamefont {Biswas}}]{Hadfield2019}%
  \BibitemOpen
  \bibfield  {author} {\bibinfo {author} {\bibfnamefont {S.}~\bibnamefont {Hadfield}}, \bibinfo {author} {\bibfnamefont {Z.}~\bibnamefont {Wang}}, \bibinfo {author} {\bibfnamefont {B.}~\bibnamefont {O'Gorman}}, \bibinfo {author} {\bibfnamefont {E.}~\bibnamefont {Rieffel}}, \bibinfo {author} {\bibfnamefont {D.}~\bibnamefont {Venturelli}},\ and\ \bibinfo {author} {\bibfnamefont {R.}~\bibnamefont {Biswas}},\ }\bibfield  {title} {\bibinfo {title} {From the quantum approximate optimization algorithm to a quantum alternating operator ansatz},\ }\href {https://doi.org/10.3390/a12020034} {\bibfield  {journal} {\bibinfo  {journal} {Algorithms}\ }\textbf {\bibinfo {volume} {12}},\ \bibinfo {pages} {34} (\bibinfo {year} {2019})}\BibitemShut {NoStop}%
\bibitem [{\citenamefont {Hadfield}(2021)}]{Hadfield2021}%
  \BibitemOpen
  \bibfield  {author} {\bibinfo {author} {\bibfnamefont {S.}~\bibnamefont {Hadfield}},\ }\bibfield  {title} {\bibinfo {title} {On the representation of boolean and real functions as hamiltonians for quantum computing},\ }\href {https://doi.org/10.1145/3478519} {\bibfield  {journal} {\bibinfo  {journal} {ACM Transactions on Quantum Computing}\ }\textbf {\bibinfo {volume} {2}},\ \bibinfo {pages} {1} (\bibinfo {year} {2021})}\BibitemShut {NoStop}%
\bibitem [{\citenamefont {B{\"a}rtschi}\ and\ \citenamefont {Eidenbenz}(2020)}]{bartschi2020grover}%
  \BibitemOpen
  \bibfield  {author} {\bibinfo {author} {\bibfnamefont {A.}~\bibnamefont {B{\"a}rtschi}}\ and\ \bibinfo {author} {\bibfnamefont {S.}~\bibnamefont {Eidenbenz}},\ }\bibfield  {title} {\bibinfo {title} {Grover mixers for qaoa: Shifting complexity from mixer design to state preparation},\ }in\ \href {https://doi.org/10.1109/QCE49297.2020.00020} {\emph {\bibinfo {booktitle} {2020 IEEE International Conference on Quantum Computing and Engineering (QCE)}}}\ (\bibinfo {organization} {IEEE},\ \bibinfo {year} {2020})\ pp.\ \bibinfo {pages} {72--82}\BibitemShut {NoStop}%
\bibitem [{\citenamefont {Wang}\ \emph {et~al.}(2020)\citenamefont {Wang}, \citenamefont {Rubin}, \citenamefont {Dominy},\ and\ \citenamefont {Rieffel}}]{Wang2020}%
  \BibitemOpen
  \bibfield  {author} {\bibinfo {author} {\bibfnamefont {Z.}~\bibnamefont {Wang}}, \bibinfo {author} {\bibfnamefont {N.~C.}\ \bibnamefont {Rubin}}, \bibinfo {author} {\bibfnamefont {J.~M.}\ \bibnamefont {Dominy}},\ and\ \bibinfo {author} {\bibfnamefont {E.~G.}\ \bibnamefont {Rieffel}},\ }\bibfield  {title} {\bibinfo {title} {$xy$ mixers: Analytical and numerical results for the quantum alternating operator ansatz},\ }\href {https://doi.org/10.1103/PhysRevA.101.012320} {\bibfield  {journal} {\bibinfo  {journal} {Phys. Rev. A}\ }\textbf {\bibinfo {volume} {101}},\ \bibinfo {pages} {012320} (\bibinfo {year} {2020})}\BibitemShut {NoStop}%
\bibitem [{\citenamefont {Fuchs}\ \emph {et~al.}(2022)\citenamefont {Fuchs}, \citenamefont {Lye}, \citenamefont {Møll~Nilsen}, \citenamefont {Stasik},\ and\ \citenamefont {Sartor}}]{fuchs2022constrained}%
  \BibitemOpen
  \bibfield  {author} {\bibinfo {author} {\bibfnamefont {F.~G.}\ \bibnamefont {Fuchs}}, \bibinfo {author} {\bibfnamefont {K.~O.}\ \bibnamefont {Lye}}, \bibinfo {author} {\bibfnamefont {H.}~\bibnamefont {Møll~Nilsen}}, \bibinfo {author} {\bibfnamefont {A.~J.}\ \bibnamefont {Stasik}},\ and\ \bibinfo {author} {\bibfnamefont {G.}~\bibnamefont {Sartor}},\ }\bibfield  {title} {\bibinfo {title} {Constraint preserving mixers for the quantum approximate optimization algorithm},\ }\href {https://doi.org/10.3390/a15060202} {\bibfield  {journal} {\bibinfo  {journal} {Algorithms}\ }\textbf {\bibinfo {volume} {15}},\ \bibinfo {pages} {202} (\bibinfo {year} {2022})}\BibitemShut {NoStop}%
\bibitem [{\citenamefont {Zhu}\ \emph {et~al.}(2022)\citenamefont {Zhu}, \citenamefont {Tang}, \citenamefont {Barron}, \citenamefont {Calderon-Vargas}, \citenamefont {Mayhall}, \citenamefont {Barnes},\ and\ \citenamefont {Economou}}]{zhu2022adaptive}%
  \BibitemOpen
  \bibfield  {author} {\bibinfo {author} {\bibfnamefont {L.}~\bibnamefont {Zhu}}, \bibinfo {author} {\bibfnamefont {H.~L.}\ \bibnamefont {Tang}}, \bibinfo {author} {\bibfnamefont {G.~S.}\ \bibnamefont {Barron}}, \bibinfo {author} {\bibfnamefont {F.~A.}\ \bibnamefont {Calderon-Vargas}}, \bibinfo {author} {\bibfnamefont {N.~J.}\ \bibnamefont {Mayhall}}, \bibinfo {author} {\bibfnamefont {E.}~\bibnamefont {Barnes}},\ and\ \bibinfo {author} {\bibfnamefont {S.~E.}\ \bibnamefont {Economou}},\ }\bibfield  {title} {\bibinfo {title} {Adaptive quantum approximate optimization algorithm for solving combinatorial problems on a quantum computer},\ }\href {https://doi.org/10.1103/PhysRevResearch.4.033029} {\bibfield  {journal} {\bibinfo  {journal} {Phys. Rev. Res.}\ }\textbf {\bibinfo {volume} {4}},\ \bibinfo {pages} {033029} (\bibinfo {year} {2022})}\BibitemShut {NoStop}%
\bibitem [{\citenamefont {Crooks}(2018)}]{crooks2018performance}%
  \BibitemOpen
  \bibfield  {author} {\bibinfo {author} {\bibfnamefont {G.~E.}\ \bibnamefont {Crooks}},\ }\href@noop {} {\bibinfo {title} {Performance of the quantum approximate optimization algorithm on the maximum cut problem}} (\bibinfo {year} {2018}),\ \Eprint {https://arxiv.org/abs/1811.08419} {arXiv:1811.08419} \BibitemShut {NoStop}%
\bibitem [{\citenamefont {Wang}\ \emph {et~al.}(2018)\citenamefont {Wang}, \citenamefont {Hadfield}, \citenamefont {Jiang},\ and\ \citenamefont {Rieffel}}]{Wang2018}%
  \BibitemOpen
  \bibfield  {author} {\bibinfo {author} {\bibfnamefont {Z.}~\bibnamefont {Wang}}, \bibinfo {author} {\bibfnamefont {S.}~\bibnamefont {Hadfield}}, \bibinfo {author} {\bibfnamefont {Z.}~\bibnamefont {Jiang}},\ and\ \bibinfo {author} {\bibfnamefont {E.~G.}\ \bibnamefont {Rieffel}},\ }\bibfield  {title} {\bibinfo {title} {Quantum approximate optimization algorithm for maxcut: A fermionic view},\ }\href {https://doi.org/10.1103/PhysRevA.97.022304} {\bibfield  {journal} {\bibinfo  {journal} {Phys. Rev. A}\ }\textbf {\bibinfo {volume} {97}},\ \bibinfo {pages} {022304} (\bibinfo {year} {2018})}\BibitemShut {NoStop}%
\bibitem [{\citenamefont {Wurtz}\ and\ \citenamefont {Love}(2021{\natexlab{a}})}]{Wurtz2020}%
  \BibitemOpen
  \bibfield  {author} {\bibinfo {author} {\bibfnamefont {J.}~\bibnamefont {Wurtz}}\ and\ \bibinfo {author} {\bibfnamefont {P.}~\bibnamefont {Love}},\ }\bibfield  {title} {\bibinfo {title} {Maxcut quantum approximate optimization algorithm performance guarantees for $p > 1$},\ }\href {https://doi.org/10.1103/PhysRevA.103.042612} {\bibfield  {journal} {\bibinfo  {journal} {Phys. Rev. A}\ }\textbf {\bibinfo {volume} {103}},\ \bibinfo {pages} {042612} (\bibinfo {year} {2021}{\natexlab{a}})}\BibitemShut {NoStop}%
\bibitem [{\citenamefont {Basso}\ \emph {et~al.}(2022{\natexlab{a}})\citenamefont {Basso}, \citenamefont {Farhi}, \citenamefont {Marwaha}, \citenamefont {Villalonga},\ and\ \citenamefont {Zhou}}]{Basso2021}%
  \BibitemOpen
  \bibfield  {author} {\bibinfo {author} {\bibfnamefont {J.}~\bibnamefont {Basso}}, \bibinfo {author} {\bibfnamefont {E.}~\bibnamefont {Farhi}}, \bibinfo {author} {\bibfnamefont {K.}~\bibnamefont {Marwaha}}, \bibinfo {author} {\bibfnamefont {B.}~\bibnamefont {Villalonga}},\ and\ \bibinfo {author} {\bibfnamefont {L.}~\bibnamefont {Zhou}},\ }\bibfield  {title} {\bibinfo {title} {The quantum approximate optimization algorithm at high depth for maxcut on large-girth regular graphs and the sherrington-kirkpatrick model},\ }in\ \href {https://doi.org/10.4230/LIPIcs.TQC.2022.7} {\emph {\bibinfo {booktitle} {17th Conference on the Theory of Quantum Computation, Communication and Cryptography (TQC 2022)}}},\ Vol.\ \bibinfo {volume} {232}\ (\bibinfo {year} {2022})\ pp.\ \bibinfo {pages} {7:1--7:21}\BibitemShut {NoStop}%
\bibitem [{\citenamefont {Marwaha}(2021)}]{Marwaha2021}%
  \BibitemOpen
  \bibfield  {author} {\bibinfo {author} {\bibfnamefont {K.}~\bibnamefont {Marwaha}},\ }\bibfield  {title} {\bibinfo {title} {Local classical max-cut algorithm outperforms $p=2$ qaoa on high-girth regular graphs},\ }\href {https://doi.org/10.22331/q-2021-04-20-437} {\bibfield  {journal} {\bibinfo  {journal} {Quantum}\ }\textbf {\bibinfo {volume} {5}},\ \bibinfo {pages} {437} (\bibinfo {year} {2021})}\BibitemShut {NoStop}%
\bibitem [{\citenamefont {Basso}\ \emph {et~al.}(2022{\natexlab{b}})\citenamefont {Basso}, \citenamefont {Gamarnik}, \citenamefont {Mei},\ and\ \citenamefont {Zhou}}]{Basso2022}%
  \BibitemOpen
  \bibfield  {author} {\bibinfo {author} {\bibfnamefont {J.}~\bibnamefont {Basso}}, \bibinfo {author} {\bibfnamefont {D.}~\bibnamefont {Gamarnik}}, \bibinfo {author} {\bibfnamefont {S.}~\bibnamefont {Mei}},\ and\ \bibinfo {author} {\bibfnamefont {L.}~\bibnamefont {Zhou}},\ }\bibfield  {title} {\bibinfo {title} {Performance and limitations of the qaoa at constant levels on large sparse hypergraphs and spin glass models},\ }in\ \href {https://doi.org/10.1109/FOCS54457.2022.00039} {\emph {\bibinfo {booktitle} {2022 IEEE 63rd Annual Symposium on Foundations of Computer Science (FOCS)}}}\ (\bibinfo {year} {2022})\ pp.\ \bibinfo {pages} {335--343}\BibitemShut {NoStop}%
\bibitem [{\citenamefont {Boulebnane}\ and\ \citenamefont {Montanaro}(2022)}]{Boulebnane2022}%
  \BibitemOpen
  \bibfield  {author} {\bibinfo {author} {\bibfnamefont {S.}~\bibnamefont {Boulebnane}}\ and\ \bibinfo {author} {\bibfnamefont {A.}~\bibnamefont {Montanaro}},\ }\href@noop {} {\bibinfo {title} {Solving boolean satisfiability problems with the quantum approximate ptimization algorithm}} (\bibinfo {year} {2022}),\ \Eprint {https://arxiv.org/abs/2208.06909} {arXiv:2208.06909} \BibitemShut {NoStop}%
\bibitem [{\citenamefont {Hadfield}\ \emph {et~al.}(2022)\citenamefont {Hadfield}, \citenamefont {Hogg},\ and\ \citenamefont {Rieffel}}]{Hadfield2022}%
  \BibitemOpen
  \bibfield  {author} {\bibinfo {author} {\bibfnamefont {S.}~\bibnamefont {Hadfield}}, \bibinfo {author} {\bibfnamefont {T.}~\bibnamefont {Hogg}},\ and\ \bibinfo {author} {\bibfnamefont {E.~G.}\ \bibnamefont {Rieffel}},\ }\bibfield  {title} {\bibinfo {title} {Analytical framework for quantum alternating operator ansätze},\ }\href {https://doi.org/10.1088/2058-9565/ACA3CE} {\bibfield  {journal} {\bibinfo  {journal} {Quantum Science and Technology}\ }\textbf {\bibinfo {volume} {8}},\ \bibinfo {pages} {015017} (\bibinfo {year} {2022})}\BibitemShut {NoStop}%
\bibitem [{\citenamefont {Herrman}(2022)}]{herrman2022relating}%
  \BibitemOpen
  \bibfield  {author} {\bibinfo {author} {\bibfnamefont {R.}~\bibnamefont {Herrman}},\ }\href@noop {} {\bibinfo {title} {Relating the multi-angle quantum approximate optimization algorithm and continuous-time quantum walks on dynamic graphs}} (\bibinfo {year} {2022}),\ \Eprint {https://arxiv.org/abs/2209.00415} {arXiv:2209.00415} \BibitemShut {NoStop}%
\bibitem [{\citenamefont {Ozaeta}\ \emph {et~al.}(2022)\citenamefont {Ozaeta}, \citenamefont {van Dam},\ and\ \citenamefont {McMahon}}]{Ozaeta2022}%
  \BibitemOpen
  \bibfield  {author} {\bibinfo {author} {\bibfnamefont {A.}~\bibnamefont {Ozaeta}}, \bibinfo {author} {\bibfnamefont {W.}~\bibnamefont {van Dam}},\ and\ \bibinfo {author} {\bibfnamefont {P.~L.}\ \bibnamefont {McMahon}},\ }\bibfield  {title} {\bibinfo {title} {Expectation values from the single-layer quantum approximate optimization algorithm on ising problems},\ }\href {https://doi.org/10.1088/2058-9565/ac9013} {\bibfield  {journal} {\bibinfo  {journal} {Quantum Science and Technology}\ }\textbf {\bibinfo {volume} {7}},\ \bibinfo {pages} {045036} (\bibinfo {year} {2022})}\BibitemShut {NoStop}%
\bibitem [{\citenamefont {Lykov}\ \emph {et~al.}(2023)\citenamefont {Lykov}, \citenamefont {Wurtz}, \citenamefont {Poole}, \citenamefont {Saffman}, \citenamefont {Noel},\ and\ \citenamefont {Alexeev}}]{Lykov2023}%
  \BibitemOpen
  \bibfield  {author} {\bibinfo {author} {\bibfnamefont {D.}~\bibnamefont {Lykov}}, \bibinfo {author} {\bibfnamefont {J.}~\bibnamefont {Wurtz}}, \bibinfo {author} {\bibfnamefont {C.}~\bibnamefont {Poole}}, \bibinfo {author} {\bibfnamefont {M.}~\bibnamefont {Saffman}}, \bibinfo {author} {\bibfnamefont {T.}~\bibnamefont {Noel}},\ and\ \bibinfo {author} {\bibfnamefont {Y.}~\bibnamefont {Alexeev}},\ }\bibfield  {title} {\bibinfo {title} {Sampling frequency thresholds for the quantum advantage of the quantum approximate optimization algorithm},\ }\href {https://doi.org/10.1038/s41534-023-00718-4} {\bibfield  {journal} {\bibinfo  {journal} {npj Quantum Information}\ }\textbf {\bibinfo {volume} {9}},\ \bibinfo {pages} {73} (\bibinfo {year} {2023})}\BibitemShut {NoStop}%
\bibitem [{\citenamefont {Shaydulin}\ \emph {et~al.}(2023)\citenamefont {Shaydulin}, \citenamefont {Li}, \citenamefont {Chakrabarti}, \citenamefont {DeCross}, \citenamefont {Herman}, \citenamefont {Kumar}, \citenamefont {Larson}, \citenamefont {Lykov}, \citenamefont {Minssen}, \citenamefont {Sun}, \citenamefont {Alexeev}, \citenamefont {Dreiling}, \citenamefont {Gaebler}, \citenamefont {Gatterman}, \citenamefont {Gerber}, \citenamefont {Gilmore}, \citenamefont {Gresh}, \citenamefont {Hewitt}, \citenamefont {Horst}, \citenamefont {Hu}, \citenamefont {Johansen}, \citenamefont {Matheny}, \citenamefont {Mengle}, \citenamefont {Mills}, \citenamefont {Moses}, \citenamefont {Neyenhuis}, \citenamefont {Siegfried}, \citenamefont {Yalovetzky},\ and\ \citenamefont {Pistoia}}]{Shaydulin2023}%
  \BibitemOpen
  \bibfield  {author} {\bibinfo {author} {\bibfnamefont {R.}~\bibnamefont {Shaydulin}}, \bibinfo {author} {\bibfnamefont {C.}~\bibnamefont {Li}}, \bibinfo {author} {\bibfnamefont {S.}~\bibnamefont {Chakrabarti}}, \bibinfo {author} {\bibfnamefont {M.}~\bibnamefont {DeCross}}, \bibinfo {author} {\bibfnamefont {D.}~\bibnamefont {Herman}}, \bibinfo {author} {\bibfnamefont {N.}~\bibnamefont {Kumar}}, \bibinfo {author} {\bibfnamefont {J.}~\bibnamefont {Larson}}, \bibinfo {author} {\bibfnamefont {D.}~\bibnamefont {Lykov}}, \bibinfo {author} {\bibfnamefont {P.}~\bibnamefont {Minssen}}, \bibinfo {author} {\bibfnamefont {Y.}~\bibnamefont {Sun}}, \bibinfo {author} {\bibfnamefont {Y.}~\bibnamefont {Alexeev}}, \bibinfo {author} {\bibfnamefont {J.~M.}\ \bibnamefont {Dreiling}}, \bibinfo {author} {\bibfnamefont {J.~P.}\ \bibnamefont {Gaebler}}, \bibinfo {author} {\bibfnamefont {T.~M.}\ \bibnamefont {Gatterman}}, \bibinfo {author} {\bibfnamefont {J.~A.}\ \bibnamefont {Gerber}}, \bibinfo {author} {\bibfnamefont
  {K.}~\bibnamefont {Gilmore}}, \bibinfo {author} {\bibfnamefont {D.}~\bibnamefont {Gresh}}, \bibinfo {author} {\bibfnamefont {N.}~\bibnamefont {Hewitt}}, \bibinfo {author} {\bibfnamefont {C.~V.}\ \bibnamefont {Horst}}, \bibinfo {author} {\bibfnamefont {S.}~\bibnamefont {Hu}}, \bibinfo {author} {\bibfnamefont {J.}~\bibnamefont {Johansen}}, \bibinfo {author} {\bibfnamefont {M.}~\bibnamefont {Matheny}}, \bibinfo {author} {\bibfnamefont {T.}~\bibnamefont {Mengle}}, \bibinfo {author} {\bibfnamefont {M.}~\bibnamefont {Mills}}, \bibinfo {author} {\bibfnamefont {S.~A.}\ \bibnamefont {Moses}}, \bibinfo {author} {\bibfnamefont {B.}~\bibnamefont {Neyenhuis}}, \bibinfo {author} {\bibfnamefont {P.}~\bibnamefont {Siegfried}}, \bibinfo {author} {\bibfnamefont {R.}~\bibnamefont {Yalovetzky}},\ and\ \bibinfo {author} {\bibfnamefont {M.}~\bibnamefont {Pistoia}},\ }\href@noop {} {\bibinfo {title} {Evidence of scaling advantage for the quantum approximate optimization algorithm on a classically intractable problem}} (\bibinfo
  {year} {2023}),\ \Eprint {https://arxiv.org/abs/2308.02342} {arXiv:2308.02342} \BibitemShut {NoStop}%
\bibitem [{\citenamefont {Stechly}\ \emph {et~al.}(2023)\citenamefont {Stechly}, \citenamefont {Gao}, \citenamefont {Yogendran}, \citenamefont {Fontana},\ and\ \citenamefont {Rudolph}}]{Stechly2023}%
  \BibitemOpen
  \bibfield  {author} {\bibinfo {author} {\bibfnamefont {M.}~\bibnamefont {Stechly}}, \bibinfo {author} {\bibfnamefont {L.}~\bibnamefont {Gao}}, \bibinfo {author} {\bibfnamefont {B.}~\bibnamefont {Yogendran}}, \bibinfo {author} {\bibfnamefont {E.}~\bibnamefont {Fontana}},\ and\ \bibinfo {author} {\bibfnamefont {M.}~\bibnamefont {Rudolph}},\ }\href@noop {} {\bibinfo {title} {Connecting the hamiltonian structure to the qaoa energy and fourier landscape structure}} (\bibinfo {year} {2023}),\ \Eprint {https://arxiv.org/abs/2305.13594} {arXiv:2305.13594} \BibitemShut {NoStop}%
\bibitem [{\citenamefont {Sun}\ \emph {et~al.}(2021)\citenamefont {Sun}, \citenamefont {Yuan}, \citenamefont {Tsunoda}, \citenamefont {Vedral}, \citenamefont {Benjamin},\ and\ \citenamefont {Endo}}]{Sun2021}%
  \BibitemOpen
  \bibfield  {author} {\bibinfo {author} {\bibfnamefont {J.}~\bibnamefont {Sun}}, \bibinfo {author} {\bibfnamefont {X.}~\bibnamefont {Yuan}}, \bibinfo {author} {\bibfnamefont {T.}~\bibnamefont {Tsunoda}}, \bibinfo {author} {\bibfnamefont {V.}~\bibnamefont {Vedral}}, \bibinfo {author} {\bibfnamefont {S.~C.}\ \bibnamefont {Benjamin}},\ and\ \bibinfo {author} {\bibfnamefont {S.}~\bibnamefont {Endo}},\ }\bibfield  {title} {\bibinfo {title} {Mitigating realistic noise in practical noisy intermediate-scale quantum devices},\ }\href {https://doi.org/10.1103/PhysRevApplied.15.034026} {\bibfield  {journal} {\bibinfo  {journal} {Phys. Rev. Appl.}\ }\textbf {\bibinfo {volume} {15}},\ \bibinfo {pages} {034026} (\bibinfo {year} {2021})}\BibitemShut {NoStop}%
\bibitem [{\citenamefont {Dasgupta}\ and\ \citenamefont {Humble}(2021)}]{Dasgupta2021}%
  \BibitemOpen
  \bibfield  {author} {\bibinfo {author} {\bibfnamefont {S.}~\bibnamefont {Dasgupta}}\ and\ \bibinfo {author} {\bibfnamefont {T.~S.}\ \bibnamefont {Humble}},\ }\href@noop {} {\bibinfo {title} {Stability of noisy quantum computing devices}} (\bibinfo {year} {2021}),\ \Eprint {https://arxiv.org/abs/2105.09472} {arXiv:2105.09472} \BibitemShut {NoStop}%
\bibitem [{\citenamefont {Herrman}\ \emph {et~al.}(2021)\citenamefont {Herrman}, \citenamefont {Ostrowski}, \citenamefont {Humble},\ and\ \citenamefont {Siopsis}}]{herrman2021lower}%
  \BibitemOpen
  \bibfield  {author} {\bibinfo {author} {\bibfnamefont {R.}~\bibnamefont {Herrman}}, \bibinfo {author} {\bibfnamefont {J.}~\bibnamefont {Ostrowski}}, \bibinfo {author} {\bibfnamefont {T.~S.}\ \bibnamefont {Humble}},\ and\ \bibinfo {author} {\bibfnamefont {G.}~\bibnamefont {Siopsis}},\ }\bibfield  {title} {\bibinfo {title} {Lower bounds on circuit depth of the quantum approximate optimization algorithm},\ }\href {https://doi.org/https://doi.org/10.1007/s11128-021-03001-7} {\bibfield  {journal} {\bibinfo  {journal} {Quantum Information Processing}\ }\textbf {\bibinfo {volume} {20}},\ \bibinfo {pages} {1} (\bibinfo {year} {2021})}\BibitemShut {NoStop}%
\bibitem [{\citenamefont {Zhou}\ \emph {et~al.}(2023)\citenamefont {Zhou}, \citenamefont {Du}, \citenamefont {Tian},\ and\ \citenamefont {Tao}}]{zhou2023qaoa}%
  \BibitemOpen
  \bibfield  {author} {\bibinfo {author} {\bibfnamefont {Z.}~\bibnamefont {Zhou}}, \bibinfo {author} {\bibfnamefont {Y.}~\bibnamefont {Du}}, \bibinfo {author} {\bibfnamefont {X.}~\bibnamefont {Tian}},\ and\ \bibinfo {author} {\bibfnamefont {D.}~\bibnamefont {Tao}},\ }\bibfield  {title} {\bibinfo {title} {Qaoa-in-qaoa: Solving large-scale maxcut problems on small quantum machines},\ }\href {https://doi.org/10.1103/PhysRevApplied.19.024027} {\bibfield  {journal} {\bibinfo  {journal} {Phys. Rev. Appl.}\ }\textbf {\bibinfo {volume} {19}},\ \bibinfo {pages} {024027} (\bibinfo {year} {2023})}\BibitemShut {NoStop}%
\bibitem [{\citenamefont {Ponce}\ \emph {et~al.}(2023)\citenamefont {Ponce}, \citenamefont {Herrman}, \citenamefont {Lotshaw}, \citenamefont {Powers}, \citenamefont {Siopsis}, \citenamefont {Humble},\ and\ \citenamefont {Ostrowski}}]{ponce2023graph}%
  \BibitemOpen
  \bibfield  {author} {\bibinfo {author} {\bibfnamefont {M.}~\bibnamefont {Ponce}}, \bibinfo {author} {\bibfnamefont {R.}~\bibnamefont {Herrman}}, \bibinfo {author} {\bibfnamefont {P.~C.}\ \bibnamefont {Lotshaw}}, \bibinfo {author} {\bibfnamefont {S.}~\bibnamefont {Powers}}, \bibinfo {author} {\bibfnamefont {G.}~\bibnamefont {Siopsis}}, \bibinfo {author} {\bibfnamefont {T.}~\bibnamefont {Humble}},\ and\ \bibinfo {author} {\bibfnamefont {J.}~\bibnamefont {Ostrowski}},\ }\href@noop {} {\bibinfo {title} {Graph decomposition techniques for solving combinatorial optimization problems with variational quantum algorithms}} (\bibinfo {year} {2023}),\ \Eprint {https://arxiv.org/abs/2306.00494} {arXiv:2306.00494} \BibitemShut {NoStop}%
\bibitem [{\citenamefont {Li}\ \emph {et~al.}(2023)\citenamefont {Li}, \citenamefont {Alam},\ and\ \citenamefont {Ghosh}}]{li2022large}%
  \BibitemOpen
  \bibfield  {author} {\bibinfo {author} {\bibfnamefont {J.}~\bibnamefont {Li}}, \bibinfo {author} {\bibfnamefont {M.}~\bibnamefont {Alam}},\ and\ \bibinfo {author} {\bibfnamefont {S.}~\bibnamefont {Ghosh}},\ }\bibfield  {title} {\bibinfo {title} {Large-scale quantum approximate optimization via divide-and-conquer},\ }\href {https://doi.org/10.1109/TCAD.2022.3212196} {\bibfield  {journal} {\bibinfo  {journal} {IEEE Transactions on Computer-Aided Design of Integrated Circuits and Systems}\ }\textbf {\bibinfo {volume} {42}},\ \bibinfo {pages} {1852} (\bibinfo {year} {2023})}\BibitemShut {NoStop}%
\bibitem [{\citenamefont {Farhi}\ \emph {et~al.}(2017)\citenamefont {Farhi}, \citenamefont {Goldstone}, \citenamefont {Gutmann},\ and\ \citenamefont {Neven}}]{Farhi2017}%
  \BibitemOpen
  \bibfield  {author} {\bibinfo {author} {\bibfnamefont {E.}~\bibnamefont {Farhi}}, \bibinfo {author} {\bibfnamefont {J.}~\bibnamefont {Goldstone}}, \bibinfo {author} {\bibfnamefont {S.}~\bibnamefont {Gutmann}},\ and\ \bibinfo {author} {\bibfnamefont {H.}~\bibnamefont {Neven}},\ }\href@noop {} {\bibinfo {title} {Quantum algorithms for fixed qubit architectures}} (\bibinfo {year} {2017}),\ \Eprint {https://arxiv.org/abs/1703.06199} {arXiv:1703.06199} \BibitemShut {NoStop}%
\bibitem [{\citenamefont {Herrman}\ \emph {et~al.}(2022)\citenamefont {Herrman}, \citenamefont {Lotshaw}, \citenamefont {Ostrowski}, \citenamefont {Humble},\ and\ \citenamefont {Siopsis}}]{Herrman2022}%
  \BibitemOpen
  \bibfield  {author} {\bibinfo {author} {\bibfnamefont {R.}~\bibnamefont {Herrman}}, \bibinfo {author} {\bibfnamefont {P.~C.}\ \bibnamefont {Lotshaw}}, \bibinfo {author} {\bibfnamefont {J.}~\bibnamefont {Ostrowski}}, \bibinfo {author} {\bibfnamefont {T.~S.}\ \bibnamefont {Humble}},\ and\ \bibinfo {author} {\bibfnamefont {G.}~\bibnamefont {Siopsis}},\ }\bibfield  {title} {\bibinfo {title} {Multi-angle quantum approximate optimization algorithm},\ }\href {https://doi.org/10.1038/s41598-022-10555-8} {\bibfield  {journal} {\bibinfo  {journal} {Scientific Reports}\ }\textbf {\bibinfo {volume} {12}},\ \bibinfo {pages} {6781} (\bibinfo {year} {2022})}\BibitemShut {NoStop}%
\bibitem [{\citenamefont {Shi}\ \emph {et~al.}(2022)\citenamefont {Shi}, \citenamefont {Herrman}, \citenamefont {Shaydulin}, \citenamefont {Chakrabarti}, \citenamefont {Pistoia},\ and\ \citenamefont {Larson}}]{Shi2022}%
  \BibitemOpen
  \bibfield  {author} {\bibinfo {author} {\bibfnamefont {K.}~\bibnamefont {Shi}}, \bibinfo {author} {\bibfnamefont {R.}~\bibnamefont {Herrman}}, \bibinfo {author} {\bibfnamefont {R.}~\bibnamefont {Shaydulin}}, \bibinfo {author} {\bibfnamefont {S.}~\bibnamefont {Chakrabarti}}, \bibinfo {author} {\bibfnamefont {M.}~\bibnamefont {Pistoia}},\ and\ \bibinfo {author} {\bibfnamefont {J.}~\bibnamefont {Larson}},\ }\bibfield  {title} {\bibinfo {title} {Multiangle qaoa does not always need all its angles},\ }\href {https://doi.org/10.1109/SEC54971.2022.00062} {\bibfield  {journal} {\bibinfo  {journal} {2022 IEEE/ACM 7th Symposium on Edge Computing (SEC)}\ ,\ \bibinfo {pages} {414}} (\bibinfo {year} {2022})}\BibitemShut {NoStop}%
\bibitem [{\citenamefont {Vijendran}\ \emph {et~al.}(2023)\citenamefont {Vijendran}, \citenamefont {Das}, \citenamefont {Koh}, \citenamefont {Assad},\ and\ \citenamefont {Lam}}]{Vijendran2023}%
  \BibitemOpen
  \bibfield  {author} {\bibinfo {author} {\bibfnamefont {V.}~\bibnamefont {Vijendran}}, \bibinfo {author} {\bibfnamefont {A.}~\bibnamefont {Das}}, \bibinfo {author} {\bibfnamefont {D.~E.}\ \bibnamefont {Koh}}, \bibinfo {author} {\bibfnamefont {S.~M.}\ \bibnamefont {Assad}},\ and\ \bibinfo {author} {\bibfnamefont {P.~K.}\ \bibnamefont {Lam}},\ }\href@noop {} {\bibinfo {title} {An expressive ansatz for low-depth quantum optimisation}} (\bibinfo {year} {2023}),\ \Eprint {https://arxiv.org/abs/2302.04479} {arXiv:2302.04479} \BibitemShut {NoStop}%
\bibitem [{\citenamefont {Wurtz}\ and\ \citenamefont {Love}(2021{\natexlab{b}})}]{wurtz2021classically}%
  \BibitemOpen
  \bibfield  {author} {\bibinfo {author} {\bibfnamefont {J.}~\bibnamefont {Wurtz}}\ and\ \bibinfo {author} {\bibfnamefont {P.~J.}\ \bibnamefont {Love}},\ }\bibfield  {title} {\bibinfo {title} {Classically optimal variational quantum algorithms},\ }\href {https://doi.org/10.1109/TQE.2021.3122568} {\bibfield  {journal} {\bibinfo  {journal} {IEEE Transactions on Quantum Engineering}\ }\textbf {\bibinfo {volume} {2}},\ \bibinfo {pages} {1} (\bibinfo {year} {2021}{\natexlab{b}})}\BibitemShut {NoStop}%
\bibitem [{\citenamefont {Chalupnik}\ \emph {et~al.}(2022)\citenamefont {Chalupnik}, \citenamefont {Melo}, \citenamefont {Alexeev},\ and\ \citenamefont {Galda}}]{chalupnik2022augmenting}%
  \BibitemOpen
  \bibfield  {author} {\bibinfo {author} {\bibfnamefont {M.}~\bibnamefont {Chalupnik}}, \bibinfo {author} {\bibfnamefont {H.}~\bibnamefont {Melo}}, \bibinfo {author} {\bibfnamefont {Y.}~\bibnamefont {Alexeev}},\ and\ \bibinfo {author} {\bibfnamefont {A.}~\bibnamefont {Galda}},\ }\bibfield  {title} {\bibinfo {title} {Augmenting qaoa ansatz with multiparameter problem-independent layer},\ }in\ \href {https://doi.org/10.1109/QCE53715.2022.00028} {\emph {\bibinfo {booktitle} {2022 IEEE International Conference on Quantum Computing and Engineering (QCE)}}}\ (\bibinfo {year} {2022})\ pp.\ \bibinfo {pages} {97--103}\BibitemShut {NoStop}%
\bibitem [{\citenamefont {Farhi}\ \emph {et~al.}(2020)\citenamefont {Farhi}, \citenamefont {Gamarnik},\ and\ \citenamefont {Gutmann}}]{farhi2020quantum}%
  \BibitemOpen
  \bibfield  {author} {\bibinfo {author} {\bibfnamefont {E.}~\bibnamefont {Farhi}}, \bibinfo {author} {\bibfnamefont {D.}~\bibnamefont {Gamarnik}},\ and\ \bibinfo {author} {\bibfnamefont {S.}~\bibnamefont {Gutmann}},\ }\href@noop {} {\bibinfo {title} {The quantum approximate optimization algorithm needs to see the whole graph: A typical case}} (\bibinfo {year} {2020}),\ \Eprint {https://arxiv.org/abs/2004.09002} {arXiv:2004.09002} \BibitemShut {NoStop}%
\bibitem [{maq()}]{maqaoarepo}%
  \BibitemOpen
  \href@noop {} {}\bibinfo {howpublished} {\url{https://github.com/GaidaiIgor/MA-QAOA}}\BibitemShut {NoStop}%
\bibitem [{\citenamefont {Hadfield}(2018)}]{hadfield2018quantum}%
  \BibitemOpen
  \bibfield  {author} {\bibinfo {author} {\bibfnamefont {S.~A.}\ \bibnamefont {Hadfield}},\ }\href {https://arxiv.org/pdf/1805.03265.pdf} {\emph {\bibinfo {title} {Quantum algorithms for scientific computing and approximate optimization}}}\ (\bibinfo  {publisher} {Columbia University},\ \bibinfo {year} {2018})\BibitemShut {NoStop}%
\bibitem [{\citenamefont {Lee}\ \emph {et~al.}(2021)\citenamefont {Lee}, \citenamefont {Saito}, \citenamefont {Cai},\ and\ \citenamefont {Asai}}]{lee2021parameters}%
  \BibitemOpen
  \bibfield  {author} {\bibinfo {author} {\bibfnamefont {X.}~\bibnamefont {Lee}}, \bibinfo {author} {\bibfnamefont {Y.}~\bibnamefont {Saito}}, \bibinfo {author} {\bibfnamefont {D.}~\bibnamefont {Cai}},\ and\ \bibinfo {author} {\bibfnamefont {N.}~\bibnamefont {Asai}},\ }\bibfield  {title} {\bibinfo {title} {Parameters fixing strategy for quantum approximate optimization algorithm},\ }in\ \href {https://doi.org/10.1109/QCE52317.2021.00016} {\emph {\bibinfo {booktitle} {2021 IEEE International Conference on Quantum Computing and Engineering (QCE)}}}\ (\bibinfo {year} {2021})\ pp.\ \bibinfo {pages} {10--16}\BibitemShut {NoStop}%
\bibitem [{\citenamefont {Wang}\ \emph {et~al.}(2021)\citenamefont {Wang}, \citenamefont {Fontana}, \citenamefont {Cerezo}, \citenamefont {Sharma}, \citenamefont {Sone}, \citenamefont {Cincio},\ and\ \citenamefont {Coles}}]{wang2021noise}%
  \BibitemOpen
  \bibfield  {author} {\bibinfo {author} {\bibfnamefont {S.}~\bibnamefont {Wang}}, \bibinfo {author} {\bibfnamefont {E.}~\bibnamefont {Fontana}}, \bibinfo {author} {\bibfnamefont {M.}~\bibnamefont {Cerezo}}, \bibinfo {author} {\bibfnamefont {K.}~\bibnamefont {Sharma}}, \bibinfo {author} {\bibfnamefont {A.}~\bibnamefont {Sone}}, \bibinfo {author} {\bibfnamefont {L.}~\bibnamefont {Cincio}},\ and\ \bibinfo {author} {\bibfnamefont {P.~J.}\ \bibnamefont {Coles}},\ }\bibfield  {title} {\bibinfo {title} {Noise-induced barren plateaus in variational quantum algorithms},\ }\href {https://doi.org/https://doi.org/10.1038/s41467-021-27045-6} {\bibfield  {journal} {\bibinfo  {journal} {Nature communications}\ }\textbf {\bibinfo {volume} {12}},\ \bibinfo {pages} {6961} (\bibinfo {year} {2021})}\BibitemShut {NoStop}%
\bibitem [{\citenamefont {Zhou}\ \emph {et~al.}(2020)\citenamefont {Zhou}, \citenamefont {Wang}, \citenamefont {Choi}, \citenamefont {Pichler},\ and\ \citenamefont {Lukin}}]{Zhou2020}%
  \BibitemOpen
  \bibfield  {author} {\bibinfo {author} {\bibfnamefont {L.}~\bibnamefont {Zhou}}, \bibinfo {author} {\bibfnamefont {S.-T.}\ \bibnamefont {Wang}}, \bibinfo {author} {\bibfnamefont {S.}~\bibnamefont {Choi}}, \bibinfo {author} {\bibfnamefont {H.}~\bibnamefont {Pichler}},\ and\ \bibinfo {author} {\bibfnamefont {M.~D.}\ \bibnamefont {Lukin}},\ }\bibfield  {title} {\bibinfo {title} {Quantum approximate optimization algorithm: Performance, mechanism, and implementation on near-term devices},\ }\href {https://doi.org/10.1103/PhysRevX.10.021067} {\bibfield  {journal} {\bibinfo  {journal} {Physical Review X}\ }\textbf {\bibinfo {volume} {10}},\ \bibinfo {pages} {021067} (\bibinfo {year} {2020})}\BibitemShut {NoStop}%
\bibitem [{\citenamefont {Boulebnane}\ and\ \citenamefont {Montanaro}(2021)}]{Boulebnane2021}%
  \BibitemOpen
  \bibfield  {author} {\bibinfo {author} {\bibfnamefont {S.}~\bibnamefont {Boulebnane}}\ and\ \bibinfo {author} {\bibfnamefont {A.}~\bibnamefont {Montanaro}},\ }\href@noop {} {\bibinfo {title} {Predicting parameters for the quantum approximate optimization algorithm for max-cut from the infinite-size limit}} (\bibinfo {year} {2021}),\ \Eprint {https://arxiv.org/abs/2110.10685} {arXiv:2110.10685} \BibitemShut {NoStop}%
\bibitem [{\citenamefont {Galda}\ \emph {et~al.}(2021)\citenamefont {Galda}, \citenamefont {Liu}, \citenamefont {Lykov}, \citenamefont {Alexeev},\ and\ \citenamefont {Safro}}]{Galda2021}%
  \BibitemOpen
  \bibfield  {author} {\bibinfo {author} {\bibfnamefont {A.}~\bibnamefont {Galda}}, \bibinfo {author} {\bibfnamefont {X.}~\bibnamefont {Liu}}, \bibinfo {author} {\bibfnamefont {D.}~\bibnamefont {Lykov}}, \bibinfo {author} {\bibfnamefont {Y.}~\bibnamefont {Alexeev}},\ and\ \bibinfo {author} {\bibfnamefont {I.}~\bibnamefont {Safro}},\ }\bibfield  {title} {\bibinfo {title} {Transferability of optimal qaoa parameters between random graphs},\ }\href {https://doi.org/10.1109/QCE52317.2021.00034} {\bibfield  {journal} {\bibinfo  {journal} {2021 IEEE International Conference on Quantum Computing and Engineering (QCE)}\ ,\ \bibinfo {pages} {171}} (\bibinfo {year} {2021})}\BibitemShut {NoStop}%
\bibitem [{\citenamefont {Sack}\ and\ \citenamefont {Serbyn}(2021)}]{Sack2021}%
  \BibitemOpen
  \bibfield  {author} {\bibinfo {author} {\bibfnamefont {S.~H.}\ \bibnamefont {Sack}}\ and\ \bibinfo {author} {\bibfnamefont {M.}~\bibnamefont {Serbyn}},\ }\bibfield  {title} {\bibinfo {title} {Quantum annealing initialization of the quantum approximate optimization algorithm},\ }\href {https://doi.org/10.22331/q-2021-07-01-491} {\bibfield  {journal} {\bibinfo  {journal} {Quantum}\ }\textbf {\bibinfo {volume} {5}},\ \bibinfo {pages} {491} (\bibinfo {year} {2021})}\BibitemShut {NoStop}%
\bibitem [{\citenamefont {Farhi}\ \emph {et~al.}(2022)\citenamefont {Farhi}, \citenamefont {Goldstone}, \citenamefont {Gutmann},\ and\ \citenamefont {Zhou}}]{Farhi2022}%
  \BibitemOpen
  \bibfield  {author} {\bibinfo {author} {\bibfnamefont {E.}~\bibnamefont {Farhi}}, \bibinfo {author} {\bibfnamefont {J.}~\bibnamefont {Goldstone}}, \bibinfo {author} {\bibfnamefont {S.}~\bibnamefont {Gutmann}},\ and\ \bibinfo {author} {\bibfnamefont {L.}~\bibnamefont {Zhou}},\ }\bibfield  {title} {\bibinfo {title} {The quantum approximate optimization algorithm and the sherrington-kirkpatrick model at infinite size},\ }\href {https://doi.org/10.22331/q-2022-07-07-759} {\bibfield  {journal} {\bibinfo  {journal} {Quantum}\ }\textbf {\bibinfo {volume} {6}},\ \bibinfo {pages} {759} (\bibinfo {year} {2022})}\BibitemShut {NoStop}%
\bibitem [{\citenamefont {Sud}\ \emph {et~al.}(2022)\citenamefont {Sud}, \citenamefont {Hadfield}, \citenamefont {Rieffel}, \citenamefont {Tubman},\ and\ \citenamefont {Hogg}}]{Sud2022}%
  \BibitemOpen
  \bibfield  {author} {\bibinfo {author} {\bibfnamefont {J.}~\bibnamefont {Sud}}, \bibinfo {author} {\bibfnamefont {S.}~\bibnamefont {Hadfield}}, \bibinfo {author} {\bibfnamefont {E.}~\bibnamefont {Rieffel}}, \bibinfo {author} {\bibfnamefont {N.}~\bibnamefont {Tubman}},\ and\ \bibinfo {author} {\bibfnamefont {T.}~\bibnamefont {Hogg}},\ }\href@noop {} {\bibinfo {title} {A parameter setting heuristic for the quantum alternating operator ansatz}} (\bibinfo {year} {2022}),\ \Eprint {https://arxiv.org/abs/2211.09270} {arXiv:2211.09270} \BibitemShut {NoStop}%
\bibitem [{\citenamefont {Sack}\ \emph {et~al.}(2023)\citenamefont {Sack}, \citenamefont {Medina}, \citenamefont {Kueng},\ and\ \citenamefont {Serbyn}}]{Sack2023}%
  \BibitemOpen
  \bibfield  {author} {\bibinfo {author} {\bibfnamefont {S.~H.}\ \bibnamefont {Sack}}, \bibinfo {author} {\bibfnamefont {R.~A.}\ \bibnamefont {Medina}}, \bibinfo {author} {\bibfnamefont {R.}~\bibnamefont {Kueng}},\ and\ \bibinfo {author} {\bibfnamefont {M.}~\bibnamefont {Serbyn}},\ }\bibfield  {title} {\bibinfo {title} {Recursive greedy initialization of the quantum approximate optimization algorithm with guaranteed improvement},\ }\href {https://doi.org/10.1103/PhysRevA.107.062404} {\bibfield  {journal} {\bibinfo  {journal} {Physical Review A}\ }\textbf {\bibinfo {volume} {107}},\ \bibinfo {pages} {062404} (\bibinfo {year} {2023})}\BibitemShut {NoStop}%
\bibitem [{\citenamefont {Sureshbabu}\ \emph {et~al.}(2023)\citenamefont {Sureshbabu}, \citenamefont {Herman}, \citenamefont {Shaydulin}, \citenamefont {Basso}, \citenamefont {Chakrabarti}, \citenamefont {Sun},\ and\ \citenamefont {Pistoia}}]{sureshbabu2023parameter}%
  \BibitemOpen
  \bibfield  {author} {\bibinfo {author} {\bibfnamefont {S.~H.}\ \bibnamefont {Sureshbabu}}, \bibinfo {author} {\bibfnamefont {D.}~\bibnamefont {Herman}}, \bibinfo {author} {\bibfnamefont {R.}~\bibnamefont {Shaydulin}}, \bibinfo {author} {\bibfnamefont {J.}~\bibnamefont {Basso}}, \bibinfo {author} {\bibfnamefont {S.}~\bibnamefont {Chakrabarti}}, \bibinfo {author} {\bibfnamefont {Y.}~\bibnamefont {Sun}},\ and\ \bibinfo {author} {\bibfnamefont {M.}~\bibnamefont {Pistoia}},\ }\href@noop {} {\bibinfo {title} {Parameter setting in quantum approximate optimization of weighted problems}} (\bibinfo {year} {2023}),\ \Eprint {https://arxiv.org/abs/2305.15201} {arXiv:2305.15201} \BibitemShut {NoStop}%
\bibitem [{\citenamefont {H\r{a}stad}(2001)}]{Hastad2001}%
  \BibitemOpen
  \bibfield  {author} {\bibinfo {author} {\bibfnamefont {J.}~\bibnamefont {H\r{a}stad}},\ }\bibfield  {title} {\bibinfo {title} {Some optimal inapproximability results},\ }\href {https://doi.org/10.1145/502090.502098} {\bibfield  {journal} {\bibinfo  {journal} {J. ACM}\ }\textbf {\bibinfo {volume} {48}},\ \bibinfo {pages} {798–859} (\bibinfo {year} {2001})}\BibitemShut {NoStop}%
\bibitem [{\citenamefont {Trevisan}\ \emph {et~al.}(2000)\citenamefont {Trevisan}, \citenamefont {Sorkin}, \citenamefont {Sudan},\ and\ \citenamefont {Williamson}}]{Trevisan2000}%
  \BibitemOpen
  \bibfield  {author} {\bibinfo {author} {\bibfnamefont {L.}~\bibnamefont {Trevisan}}, \bibinfo {author} {\bibfnamefont {G.~B.}\ \bibnamefont {Sorkin}}, \bibinfo {author} {\bibfnamefont {M.}~\bibnamefont {Sudan}},\ and\ \bibinfo {author} {\bibfnamefont {D.~P.}\ \bibnamefont {Williamson}},\ }\bibfield  {title} {\bibinfo {title} {Gadgets, approximation, and linear programming},\ }\href {https://doi.org/10.1137/S0097539797328847} {\bibfield  {journal} {\bibinfo  {journal} {SIAM Journal on Computing}\ }\textbf {\bibinfo {volume} {29}},\ \bibinfo {pages} {2074} (\bibinfo {year} {2000})}\BibitemShut {NoStop}%
\end{thebibliography}%
\clearpage

\section*{Supplemental Information}

\subsection{Comparison of QAOA Relax and Random initialization strategies}

\begin{figure}[b]
    \centering
    \begin{subfigure}[b]{0.49\textwidth}
        \centering
        \includegraphics[width=\textwidth]{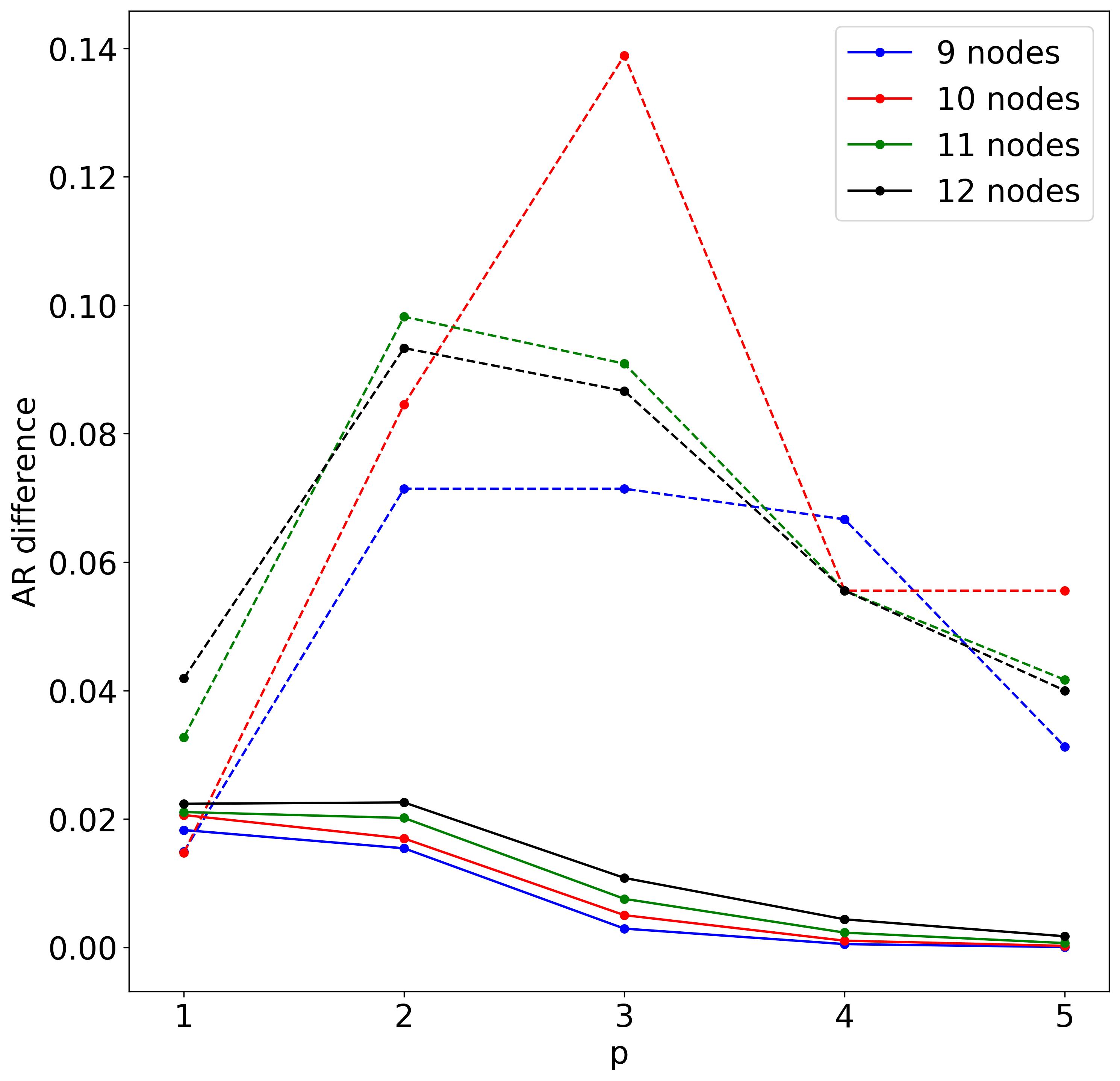}
        \caption{Fixed c-depth = 3, varying number of nodes}
    \end{subfigure}
    \hfill
    \begin{subfigure}[b]{0.49\textwidth}
        \centering
        \includegraphics[width=\textwidth]{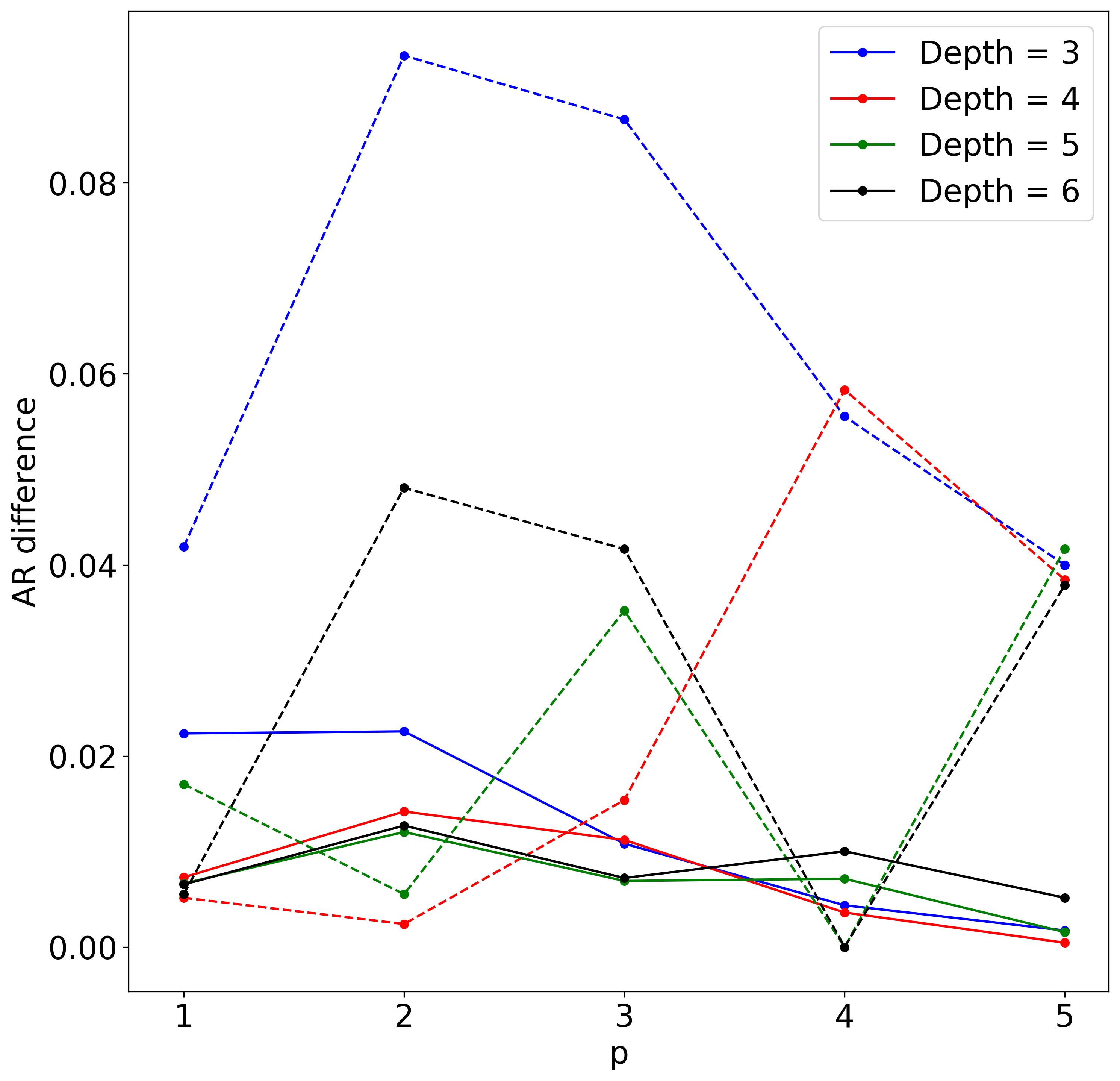}
        \caption{Fixed number of nodes = 12, varying c-depth}
    \end{subfigure}
    \caption{AR of QAOA Relax relative to AR of Random. Solid (dashed) lines show differences in average (worst case) approximation ratio.}
    \label{fig:qaoa-relax-random-diff}
\end{figure}

In the main text, we introduced a new initialization strategy for MA-QAOA, called QAOA Relax, and demonstrated its advantage over Random initialization strategy, used in the previous studies of MA-QAOA, but only on one data set. In Figure \ref{fig:qaoa-relax-random-diff}, we show how this advantage changes as we increase the number of nodes or c-depth of the graphs. As one can see, the average AR difference steadily increases for all $p$ as the number of nodes is increased, which makes this initialization strategy even more useful for larger graphs. The decrease in the difference of approximation ratios at larger values of $p$ is due to the fact that QAOA Relax is already converged by $p \in \{2, 3\}$, as can be seen in Figure \ref{fig:qaoa-ma-performance}, but Random converges later. The difference is even more pronounced when comparing the worst case ARs, achieving up to +0.14 AR on the considered data sets.

\subsection{Comparison of different values for the Constant initialization strategy}

\begin{figure}
    \centering
    \includegraphics[width=0.5\textwidth]{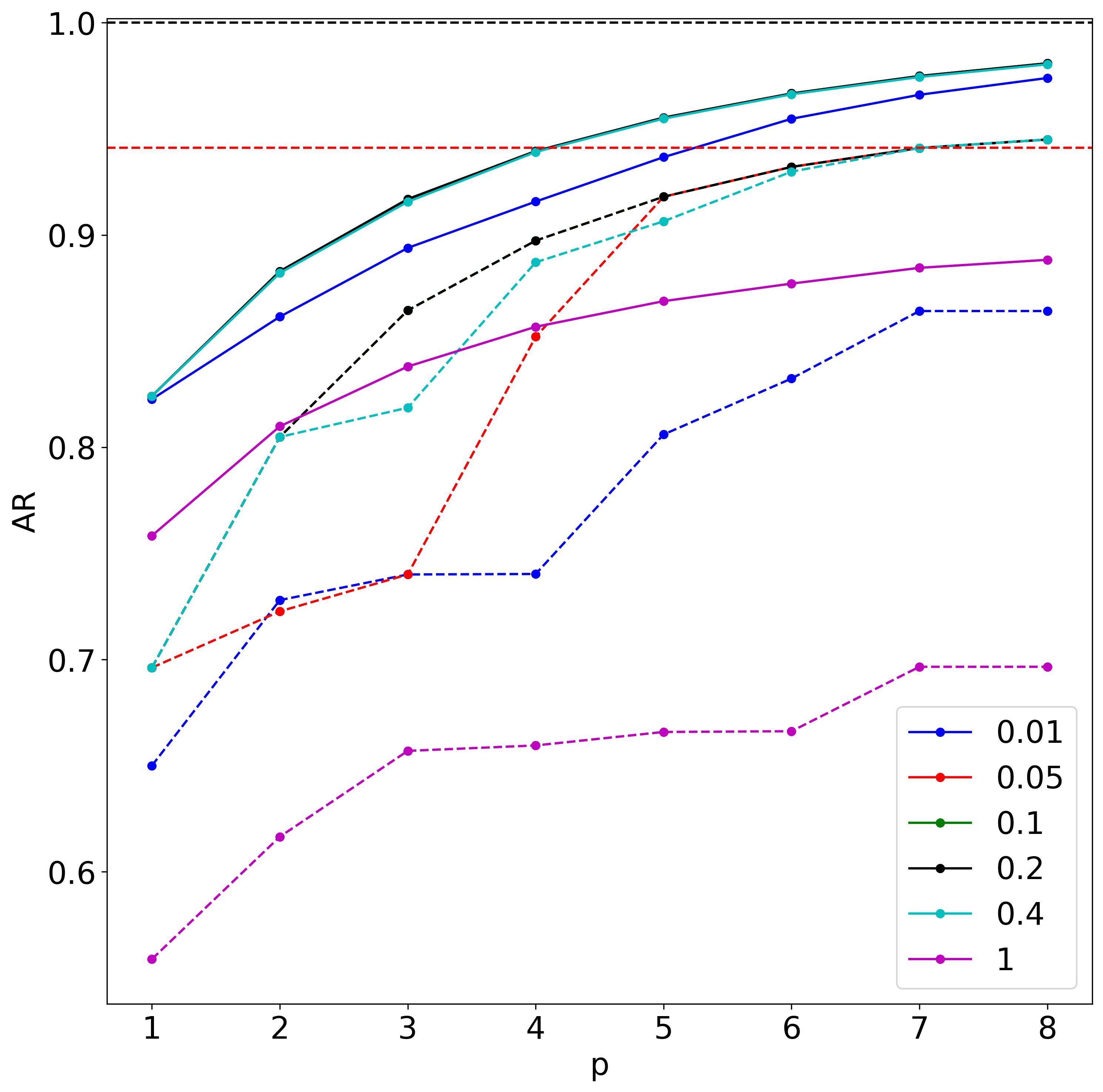}
    \caption{AR obtained with different values of the Constant initialization strategy. Solid (dashed) lines show differences in average (worst case) approximation ratio.}
    \label{fig:qaoa-constant}
\end{figure}

In the main text we mentioned that we tried a few different values for the Constant initialization strategy. In Figure \ref{fig:qaoa-constant} we show the detailed comparison between the ARs achieved with the each value on the data set with 9 nodes (same as in Figure \ref{fig:qaoa-heuristics}). The values from 0.05 to 0.4 have nearly the same AR on average, but can be distinguished in the worst case performance. The values of 0.1 and 0.2 have nearly the same worst-case performance too, but 0.2 converges somewhat faster, thus it was selected as the best value in the main text.

\subsection{QAOA heuristics as a function of cost}

\begin{figure}
    \centering
    \includegraphics[width=0.5\textwidth]{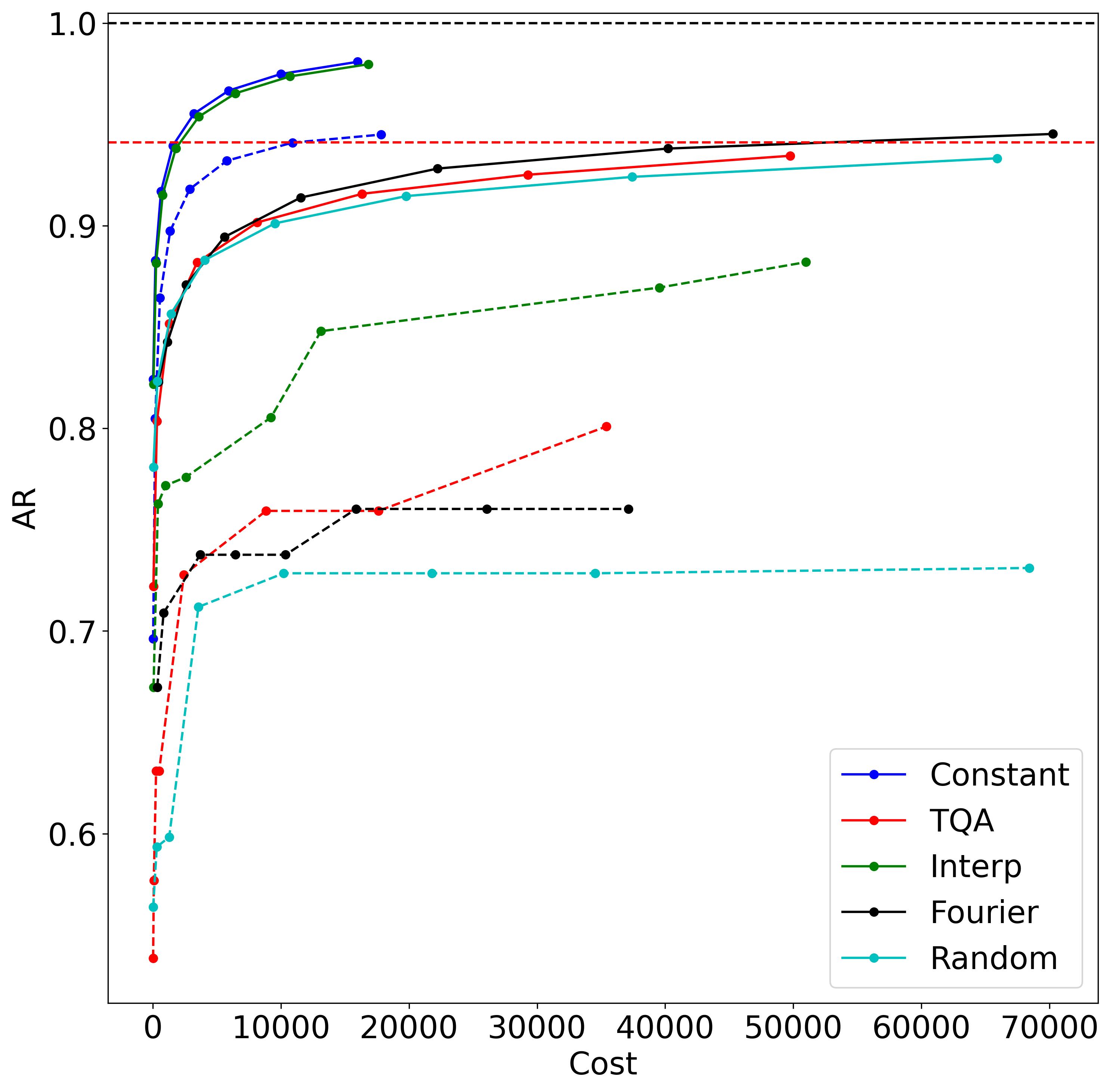}
    \caption{AR vs cost for the QAOA initialization strategies considered in Figure \ref{fig:qaoa-heuristics}a. Solid (dashed) lines show differences in average (worst case) approximation ratio.}
    \label{fig:qaoa-heuristics-cost}
\end{figure}

It is interesting to repeat the analysis of section \ref{sec:qaoa_strategies} with the cost definition of Eq. \ref{eq:cost} to take into account not only the number of layers, but also the number of QPU calls. The result of this is shown in Figure \ref{fig:qaoa-heuristics-cost}. As one can see, the Constant initialization strategy still remains a winner, especially in the worst case, achieving not only the largest AR, but also using the smallest number of calls for the optimization procedure.

\end{document}